\documentclass[%
reprint,
 amsmath,amssymb,
 aps,
 prb,
 citeautoscript,
]{revtex4-2}

\usepackage{graphicx}% Include figure files
\usepackage{dcolumn}% Align table columns on decimal point
\usepackage{bm}% bold math
\usepackage{xcolor}
\usepackage{hyperref}% add hypertext capabilities
\hypersetup{
    colorlinks=true,
    linkcolor=blue,
    citecolor=blue,
    filecolor=blue,      
    urlcolor=blue,
    }
\newcommand{\FS}[1]{\left\langle #1 \right\rangle_{\scriptscriptstyle\mathrm{FS}}}
 
\newcommand{\kb}{k_\mathrm{B}}
\newcommand{\tc}{T_\mathrm{c}}

\newcommand{\RR}{\mathbf{R}}
\newcommand{\vF}{\mathbf{v}_\mathrm{F}}
\newcommand{\pF}{\mathbf{p}_\mathrm{F}}

\newcommand{\NF}{\mathcal{N}_\mathrm{F}}

\newcommand{\xx}{\emph{\fontfamily{cmss}\selectfont x}}

\begin{document}

\title{A finite element method for the quasiclassical theory of superconductivity}

\author{Kevin Marc Seja}
\author{Tomas L\"ofwander}
\affiliation{Department of Microtechnology and Nanoscience - MC2,
Chalmers University of Technology,
SE-41296 G\"oteborg, Sweden}

\date{\today}

\begin{abstract}
The Eilenberger-Larkin-Ovchinnikov-Eliashberg quasiclassical theory of superconductivity is a powerful method enabling studies of a wide range of equilibrium and non-equilibrium phenomena in conventional and unconventional superconductors.
We introduce here a finite element method, based on a discontinuous Galerkin approach, to self-consistently solve the underlying transport equations for general device geometries, arbitrary mean free path and symmetry of the superconducting order parameter.
We present results on i) the influence of scalar impurity scattering on phase crystals in $d$-wave superconducting grains at low temperatures and ii) the current flow and focusing in $d$-wave superconducting weak links, modeling recent experimental realizations of grooved high-temperature superconducting Dayem bridges. The high adaptability of this finite element method for quasiclassical theory paves the way for future investigations of superconducting devices and new physical phenomena in unconventional superconductors.
\end{abstract}

\maketitle

\section{Introduction}

The quasiclassical theory of superconductivity \cite{eilenberger_transformation_1968,Larkin1969,Eliashberg1971} allows the study of a wide range of phenomena in both conventional and unconventional superconductors on a mesoscopic scale.
Solving the underlying transport equations is, however, a highly involved numerical problem, especially when a self-consistent determination of the self-energies is required.
We propose here a finite element method (FEM), specifically a discontinuous Galerkin (DG) approach, that can be used to solve the underlying transport equations of the full Eilenberger quasiclassical theory in realistic superconducting systems in two or three dimensions.  
The DG method of solving transport equations in dimension $D \geq 2$ was first proposed in the 1970s for the neutron transport equation in the context of nuclear reactors \cite{osti_4491151,Lasaint1974Jan,osti_4237082} but has since been applied to (in)compressible flow dynamics, chemical transport, and many other areas of physics \cite{Cockburn2003Nov}.

Such a FEM offers several advantages over finite difference methods used in the past \cite{Sauls2009,Hakansson2015Sep}. Firstly, the solution strategy introduced here is directly applicable to systems in one, two, or three dimensions, up to a slight adaption of the boundary conditions depending on the dimension. Secondly, it provides better adaptability of the discretization to the geometry under investigation as well as the degree of approximating functions which improves convergence \cite{Cockburn2000}.

For dirty conventional superconductors, where the mean free path $\ell$ is much shorter than the superconducting coherence length $\xi_0$, the diffusive approximation due to Usadel is valid. An implementation of a FEM for the underlying Usadel diffusion equations has been reported \cite{Amundsen2016Mar}, and used to analyze experimental results \cite{Lahabi2017Dec} as well as to investigate new physics in higher-dimensional structures \cite{Amundsen2017Aug}. The method we present here is valid for superconductors with arbitrary mean free path and order-parameter symmetry, and thus extends the possibility of finite-element analysis to a wide range of superconducting materials and phenomena.

This paper is organized as follows. In Sec.~\ref{sec:Theory}, we review the underlying equations of the quasiclassical theory of superconductivity and present the reformulation in terms of a discontinous Galerkin method. This is followed by \textcolor{black}{results on two example problems} in Sec.~\ref{sec:results}. Firstly, we investigate the effect of scalar impurites on phase crystals in a closed system. Secondly, we model a current biased superconducting Dayem bridge as an example for current flow and focusing in a geometry with open boundaries. We conclude with an outlook on possible further applications of the method in Sec.~\ref{sec:discussion}.

\section{Theory}
\label{sec:Theory}
\subsection{Quasiclassical theory}
The core of the quasiclassical theory of superconductivity is the Eilenberger equation,
\begin{align}
i \hbar \vF \cdot \nabla &\check{g}(\pF,\RR,\varepsilon) 
\nonumber
\\
&+ \left[ \varepsilon \hat{\tau}_3 \check{1} - \check{h}(\pF,\RR,\varepsilon), \check{g}(\pF,\RR,\varepsilon) \right] = 0,
\label{eq:transportequation}
\end{align}
for the quasiclassical Green's function $\check{g}(\pF,\RR,\varepsilon)$ together with a normalization condition
\begin{equation}
\check{g}(\pF,\RR,\varepsilon)^2 = -\pi^2.
\label{eq:normalizationCondition}
\end{equation}
The above form is for the time-independent steady state that we assume in this paper. In this case, the propagators only depend on momentum direction on the Fermi surface, $\pF$, spatial coordinate, $\RR$, and energy, $\varepsilon$. In Eq.~\eqref{eq:transportequation}, $\vF$ is the Fermi velocity, the $[A,B]$ denotes a commutator between matrices $A$ and $B$, a $\check{~}$ marks Keldysh-space matrices, $\hat{\tau}_3$ is the third Pauli matrix in Nambu (particle-hole) space, indicated by a $\hat{~}$, and $\check{h}$ is the self-energy matrix. The three elements of the Keldysh-space matrix $\check{g}$ are
\begin{equation}
\check{g}(\pF,\RR,\varepsilon) =
\begin{pmatrix}
\hat{g}^\mathrm{R}(\pF,\RR,\varepsilon) & \hat{g}^\mathrm{K}(\pF,\RR,\varepsilon)
\\
0 & \hat{g}^\mathrm{A}(\pF,\RR,\varepsilon)
\end{pmatrix},
\end{equation}
where the spectrum of the system is determined by the retarded (advanced) component $\hat{g}^R$ ($\hat{g}^A$), while information about the occupation of states is contained in the Keldysh component $\hat{g}^K$. All three components are in turn matrices in Nambu space that we will detail below. To shorten the notation, we often drop the explicit dependences on $\pF$, $\RR$, and $\varepsilon$. The selfenergy matrix $\check{h}$ has the same form as $\check{g}$ in Keldysh space with the Nambu-space components
\begin{align}
\hat{h}^{\mathrm{R}, \mathrm{A}} = 
\begin{pmatrix}
\Sigma & \Delta\\
\tilde{\Delta} & \tilde{\Sigma}
\end{pmatrix}^{\mathrm{R}, \mathrm{A}},
~
\hat{h}^\mathrm{K}
=
\begin{pmatrix}
\Sigma & \Delta
\\
-\tilde{\Delta} & -\tilde{\Sigma}
\end{pmatrix}^\mathrm{K}.
\label{eq:selfenergy_elements}
\end{align}
This set of equations was first derived by Eilenberger \cite{eilenberger_transformation_1968}, and separately Larkin and Ovchinnikov \cite{Larkin1969}, and generalized to non-equilibrium by Eliashberg \cite{Eliashberg1971}.

Instead of solving Eq.~\eqref{eq:transportequation} directly, it is advantageous to use parametrizing functions for $\check{g}$ that guarantee that the normalization condition, Eq.~\eqref{eq:normalizationCondition}, is satisfied. One well-established choice is a parametrization in terms of coherence amplitudes $\gamma, \tilde{\gamma}$ \cite{Nagato1993, Schopohl1995, Schopohl1998, Shelankov2000Mar}
%, \textcolor{black}{which is related to solutions of the Andreev equations}\cite{}, 
and generalized distribution functions $\xx, \tilde{\xx}$ \cite{eschrig_distribution_2000,Eschrig2009Oct}.
The coherence amplitudes determine the retarded component
\begin{align}
\hat{g}^\mathrm{R} = 
-2\pi i
\!
\begin{pmatrix}
\mathcal{G}& \mathcal{F}
 \\
-\tilde{\mathcal{F}} & -\tilde{\mathcal{G}}
\end{pmatrix}^\mathrm{R}
+ i \pi \hat{\tau}_3,
\end{align}
where $\mathcal{G}^R \equiv ( 1 - \gamma^\mathrm{R} \tilde{\gamma}^\mathrm{R})^{-1}$ and $\mathcal{F}^\mathrm{R} \equiv \mathcal{G}^\mathrm{R}\gamma^\mathrm{R}$.
The expression for the advanced element $\hat{g}^\mathrm{A}$ is analogous to $\hat{g}^\mathrm{R}$. The Keldysh component also involves the distribution functions $\xx$ and $\tilde{\xx}$, 
\begin{align}
\hat{g}^\mathrm{K}
&= 
\begin{pmatrix}
g& f
\\
- \tilde{f} & - \tilde{g}
\end{pmatrix}^\mathrm{K} 
\nonumber
\\
&
\equiv
-2 \pi i 
\begin{pmatrix}
\mathcal{G} & \mathcal{F}
\\
-\tilde{\mathcal{F}} & -\tilde{\mathcal{G}}
\end{pmatrix}^\mathrm{R}
\begin{pmatrix}
\xx & 0
\\ 0 & \tilde{\xx}
\end{pmatrix}
\begin{pmatrix}
\mathcal{G} & \mathcal{F}
\\
-\tilde{\mathcal{F}} & -\tilde{\mathcal{G}}
\end{pmatrix}^\mathrm{A}.
\label{eq:riccati_x}
\end{align}
Generally all elements of $\hat{g}^\mathrm{R,A,K}$ are matrices in a space of internal degrees of freedom such as spin, but in the following we assume a spin-singlet superconductor for simplicity. Then, $\gamma$ can be written as $\gamma = \gamma_\mathrm{singlet} i \sigma_2$ -- where $\gamma_\mathrm{singlet}$ is a scalar function and $\sigma_2$ is the second Pauli matrix in spin space -- while $\xx$ is proportional to the unit matrix in spin space. One central symmetry of the theory is particle-hole conjugation, expressed as
\begin{align}
\tilde{A}(\pF,\RR,\varepsilon) = A^*(-\pF,\RR,-\varepsilon^*).
\label{eq:TildeSymmetry}
\end{align}

Starting from Eq.~\eqref{eq:transportequation}, a set of coupled equations for the parametrizing functions can be derived. In the steady-state, the equations for $\gamma(\pF,\RR,\varepsilon)$ and $\xx(\pF,\RR,\varepsilon)$ read 
\begin{align}
&\left( i\hbar\vF\cdot\nabla + 2 \varepsilon \right) \gamma^{\mathrm{R},\mathrm{A}}
= \bigl( \gamma \tilde{\Delta} \gamma + \Sigma \gamma - \gamma \tilde{\Sigma} - \Delta \bigr)^{\mathrm{R},\mathrm{A}}\!,
\label{eq:GammaEquation}
\\
&i \hbar \vF \cdot \nabla \xx - \left( \gamma \tilde{\Delta} + \Sigma \right)^\mathrm{R} \xx - \xx \left( \Delta \tilde{\gamma} - \Sigma \right)^\mathrm{A}
\nonumber
\\
&\qquad \qquad=
- \gamma^\mathrm{R} \tilde{\Sigma}^\mathrm{K} \tilde{\gamma}^\mathrm{A} + \Delta^\mathrm{K} \tilde{\gamma}^\mathrm{A} + \gamma^\mathrm{R} \tilde{\Delta}^\mathrm{K} - \Sigma^\mathrm{K}.
\label{eq:XEquation}
\end{align}
Equations for the remaining amplitudes $\tilde{\gamma}^\mathrm{R/A}$ and distribution $\tilde{\xx}$ can be obtained via the tilde symmetry in Eq.~\eqref{eq:TildeSymmetry}.
Note that both equations are \textit{transport equations} along a transport direction $\vF$, indicated by the directional derivative $\vF \cdot \nabla$. Solutions of these differential equations have to be found along semi-classical trajectories determined by $\vF$, a starting point, and an end point. \textcolor{black}{Along such a trajectory, Eq.~\eqref{eq:GammaEquation} is a Riccati equation that is well-studied in literature \cite{Reid1972Jan}.}
Observables are then determined by an average over all possible momentum orientations $\vF$ on the Fermi surface. For our two-dimensional case we use a circular Fermi surface, where the orientation is fully determined by the angle to the $x$ axis, $\varphi_\mathrm{F}$. The average of an observable or self-energy $A$ is then
\begin{equation}
\FS{A(\pF)} = \frac{1}{2\pi} \int\limits_{0}^{2\pi}\!\mathrm{d}\varphi_\mathrm{F}\!~A\left(\varphi_\mathrm{F}\right)
.\label{eq:FermiSurfaceAverage}
\end{equation}
After solving Eq.~\eqref{eq:transportequation}, all self-energies have to be recalculated until self-consistency is achieved. In the present paper, we use 
\begin{equation}
\check{h}(\pF,\RR,\varepsilon) = \check{h}_\mathrm{mf}(\pF,\RR) + \check{h}_\mathrm{s}(\RR,\varepsilon),
\end{equation}
where $\check{h}_\mathrm{mf}$ is the mean-field order parameter, given by
\begin{equation}
\Delta_0(\RR) = \lambda\NF\!\int\limits_{-\varepsilon_\mathrm{c}}^{\varepsilon_c}\!\frac{d\varepsilon}{8\pi i}
\FS{\mbox{Tr}\left[i\sigma_2\eta_d(\pF) f^\mathrm{K}(\pF,\RR,\varepsilon)\right]}.
\label{eq:gapequation}
\end{equation}
Here, $\eta_d$ is the basis function for the respective order-parameter symmetry. 
%For an $s$-wave superconductor, we have $\eta_d(\varphi_\mathrm{F}) = 1$. 
For the $d_{x^2 - y^2}$ order parameter, which we will consider in this paper, we can choose $\eta_d = \cos \left[2\left(\varphi_\mathrm{F} - \alpha \right)\right]$ where the angle $\alpha$ specifies the misalignment of the main crystal axis to the geometric axis. 
The self-energy $\check{h}_\mathrm{s}$ describes scalar impurity scattering. Assuming an average dilute impurity concentration $n_\mathrm{i}$, the impurity self-energy is found from the t-matrix equation in the non-crossing approximation \cite{AGD}:
\begin{align}
\check{h}_s = n_\mathrm{i} \check{t} \equiv n_\mathrm{i}
\begin{pmatrix}
\hat{t}^R & \hat{t}^K
\\
0 & \hat{t}^A
\end{pmatrix}.
\label{eq:selfEnergyTMatrixEquation}
\end{align}
For scattering that is isotropic in momentum space with an $s$-wave scattering potential $u_0$ the elements of $\check{t}$ satisfy the equations
\begin{align}
\hat{t}^\mathrm{R,A} &= \frac{u_0 \hat{1} + u_0^2 \NF \FS{\hat{g}^\mathrm{R,A}}}{\hat{1} - \left[  u_0 \NF \FS{\hat{g}^\mathrm{R,A} }\right]^2   },
\\
\hat{t}^\mathrm{K} &= \NF \hat{t}^\mathrm{R} \FS{\hat{g}^\mathrm{K}} \hat{t}^\mathrm{A}.
\label{eq:tKDefinition}
\end{align}
The two parameters of the model, the isotropic impurity potential $u_0$ and impurity concentration $n_i$, can equivalently be written as a scattering energy $\Gamma_u=n_\mathrm{imp}/\pi\NF$ and a scattering phase shift, $\delta_0=\arctan(\pi \NF u_0)$. They can be combined into 
%\begin{align}
$\Gamma \equiv \Gamma_u \sin^2 \delta_0$,
%\label{eq:PairBreakingParameter}
%\end{align}
the so-called pair-breaking energy, which determines the normal-state mean free path
%\begin{align}
$\ell = \hbar v_\mathrm{F}/(2\Gamma)$.
%\label{eq:mfpGammaRelation}
%\end{align} 
We restrict ourselves here to the weak-scattering Born limit,
\begin{align}
\hat{h}_\mathrm{s, Born}^\mathrm{R}(\RR,\varepsilon) = \frac{\Gamma}{\pi}
\FS{\hat{g}^R(\pF,\RR,\varepsilon)},
\label{eq:ImpuritiesBorn}
\end{align}
or the opposite strong-scattering unitary limit,
\begin{align}
\hat{h}_\mathrm{s, unit.}^\mathrm{R}(\RR,\varepsilon) = 
-
\pi \Gamma
\frac{ \FS{\hat{g}^R(\pF,\RR,\varepsilon)} }{\FS{\hat{g}^R(\pF,\RR,\varepsilon)}^2}.
\label{eq:ImpuritiesUnitary}
\end{align}
\textcolor{black}{
Solving Eqs.~\eqref{eq:GammaEquation}-\eqref{eq:XEquation} allows to construct the full quasiclassical Green's function $\check{g}$ that solves Eq.~\eqref{eq:transportequation}, and can be used to calculate physical observables. For example, we find the momentum-resolved density of states
}
\begin{equation}
\mathcal{N}(\pF, \RR, \varepsilon) = -\frac{1}{4\pi}\mathrm{Im} \mathrm{Tr}\left[\hat{\tau}_3 \hat{g}^R(\pF, \RR, \varepsilon)\right],
\end{equation}
which determines the full density of states,
\begin{equation}
N(\RR,\varepsilon) = 2\NF\FS{\mathcal{N}(\pF, \RR, \varepsilon)}.
\end{equation}
Here, $\mathcal{N}_\mathrm{F}$ is the normal-state density of states per spin at the Fermi level.
The charge current density is obtained via
\begin{align}
\mathbf{j}(\RR) &= e\NF\!\int\limits_{-\infty}^{\infty}\!
\frac{\mathrm{d}\varepsilon}{8\pi i} 
\FS{\mathrm{Tr}\left[ \vF \hat{\tau}_3 \hat{g}^\mathrm{K}(\pF,\RR,\varepsilon) \right]}
\label{eq:ChargeCurrentDefinition}.
\end{align}
\textcolor{red}{We use $j_0=e v_\mathrm{F}\NF\kb\tc$ as the unit for charge current.}
If both the system and all connected reservoirs are in equilibrium the Keldysh Green's function at temperature $T$ simply reads
\begin{align}
\hat{g}^\mathrm{K} = \left( \hat{g}^\mathrm{R} -\hat{g}^\mathrm{A}\right) \tanh \frac{\varepsilon}{2 T}.
\end{align}
By replacing the energy $\varepsilon$ with imaginary-axis Matsubara frequencies $ i \varepsilon_n = i \pi \kb T(2n+1)$ in Eq.~\eqref{eq:GammaEquation} we can then solve for the Matsubara coherence function $\gamma^\mathrm{M}$. For better numerical performance alternatives such as the Ozaki poles of a continued-fraction expansion \cite{Ozaki2007Jan} can be used. This is the case for the results presented in this paper.

\subsection{Trajectory method}
\label{sec:trajectory_method}
Self-consistent numerical solutions of Eqs.~\eqref{eq:GammaEquation} and \eqref{eq:XEquation} require a discretization of the self-energies on a grid, and similarly discretizing the trajectories in ''steps``. In (quasi-) one-dimensional systems it is always possible to choose the two to be commensurate. One possible choice is to model the energy landscape and the coherence functions as piecewise constant in between gridpoints. \textcolor{black}{In this case, analytic solutions to Eq.~\eqref{eq:GammaEquation}-\eqref{eq:XEquation} in a region of constant selfenergies can be found.} This leads to a \textit{stepping method} where the solution for the equation of motion, Eq.~\eqref{eq:GammaEquation}, is found along the trajectory by stepping along the trajectory step-by-step from one region to another.

\begin{figure}
    \centering
    \includegraphics[width=7cm]{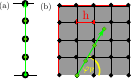}
    \caption{(a) For quasi one-dimensional models, the trajectory steps can be chosen to be commensurate with the grid of self-energy points. (b) Already in two dimensions, this is not possible. As an example, the starting part of a trajectory for angle $\varphi_\mathrm{F}$ is shown. For equal grid spacing and step size $h$ the steps (green dots) along the trajectory do not coincide with the self-energy grid (black dots).}
    \label{fig:gridProblems}
\end{figure}

In contrast, the steps cannot be made commensurate with the self-energy grid for all momentum orientations at once in two or three dimensions, as sketched Fig.~\ref{fig:gridProblems}. A stepping method is still possible and has been implemented for two-dimensional systems \cite{Hakansson2015Sep}. %SupgraCongaPaper}. 
The discrepancy between the self-energy grid and the trajectory steps require frequent interpolation from the self-energy grid to the ''stepping grid``.

Here, we describe an alternative solution strategy that relies on a finite element method, more specifically a discontinuous Galerkin method. The solution to the differential equation is then defined on the same grid as the self-energies which avoids grid interpolation.

\subsection{Discontinuous Galerkin method for the coherence amplitude $\gamma$}
\label{sec:FiniteElementSection}
\begin{figure}
    \centering
    \includegraphics[width=\columnwidth]{{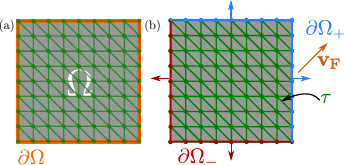}}
    \caption{(a) Example for a domain $\Omega$ (grey) with boundary $\partial \Omega$ (orange), and a triangulation (green). The green dots mark geometric nodes of the underlying mesh, while the green edges mark the outline of the triangulation cells. (b) Inflow (outflow) boundaries $\partial \Omega_-$ ($\partial \Omega_+$) in dark red (light blue) for a given transport direction $\vF$ (orange arrow). The small outer arrows indicate the (outward-pointing) surface normals. All cell edges that are not part of the boundary, marked in green, form the internal edges labelled $\tau$. }
    \label{fig:mesh_example}
\end{figure}
The underlying assumption of finite element methods is that a given geometry can be decomposed into a collection of elements or cells, as indicated in Fig.~\ref{fig:mesh_example}. Note that we use a two-dimensional geometry and triangular cells for illustration, but the method introduced here can be directly applied to three-dimensional models as well.
The original differential equation then has to be translated to its weak form. 
We will focus on the Riccati equation for $\gamma^\mathrm{R}$ but will drop the superscript $~^\mathrm{R}$ in the following. The derivation is largely identical for the advanced function $\gamma^\mathrm{A}$.
%For brevity, we omit the explicit dependence on the spatial coordinate $\mathbf{R}$, energy $\varepsilon$, and momentum orientation $\mathbf{p}_\mathrm{F}$ on all functions in the following. 
For ease of notation, we focus on the spin-degenerate case now which will result in a scalar equation for the coherence amplitude $\gamma$. The generalization to the full two-by-two spin structure follows along identical lines but will produce a coupled system of equations for four unknowns, with equally many independent test functions.

We start from Eq.~\eqref{eq:GammaEquation} that can, in the scalar case, be rearranged to read 
\begin{align}
i\hbar\vF\cdot\nabla \gamma
+ \gamma \tilde{\Delta} \gamma  + 2 \varepsilon \gamma - \Sigma \gamma + \gamma \tilde{\Sigma} \!= - \Delta .
\end{align}
Multiplying both sides by a, for now unspecified, test function $\phi({\RR})\equiv \phi$, and integrating over the entire domain $\Omega$ gives
\begin{align}
i \hbar \int\limits_\Omega
\phi ~\vF\cdot\nabla \gamma  ~\mathrm{d}\Omega
&+ \int\limits_\Omega\! \phi \left(  \gamma \tilde{\Delta} \gamma + 2  \varepsilon \gamma   - \Sigma \gamma  + \gamma \tilde{\Sigma} \right) \mathrm{d}\Omega
\nonumber
\\
&= - \int\limits_\Omega \phi \Delta ~\mathrm{d}\Omega.
\end{align}
We now assume to have a triangulation $\mathcal{T}$, meaning a collection of triangles $T_j$ that satisfy $\Omega = \cup_{T_j \in \mathcal{T}} T_j$. The integration is then split into a sum of integrals over each triangle,
\begin{align}
\sum\limits_{T_j \in \mathcal{T}}
\Biggl[ &i \hbar\!\int\limits_{T_j}\!
\phi ~\vF\cdot\nabla \gamma  ~\mathrm{d}\Omega_j
+ \int\limits_{T_j}\! \phi \left( \gamma \tilde{\Delta} \gamma \right)\mathrm{d}\Omega_j
\nonumber
\\
&+\!\int\limits_{T_j}\! \phi \left(  2  \varepsilon \gamma  - \Sigma \gamma  + \gamma \tilde{\Sigma} \right)\mathrm{d}\Omega_j \Biggr]
= - \sum\limits_{T_j \in \mathcal{T}} \int\limits_{T_j}\!\phi \Delta  ~\mathrm{d}\Omega_j.
\label{eq:WeakFormIntermediate}
\end{align}
Performing a partial integration on the first term on the left hand side in Eq.~\eqref{eq:WeakFormIntermediate}, we arrive at
\begin{align}
\sum\limits_{T_j \in \mathcal{T}}\!\Biggl[
&i \hbar\!\int\limits_{\partial \Omega_j}\!
\phi~(\gamma \vF)\cdot \mathbf{n}_j~\mathrm{d}s_j
- i \hbar \int\limits_{T_j}\!\gamma\vF\cdot (\nabla \phi)\mathrm{d}\Omega_j
\nonumber
\\
&+ \int\limits_{T_j}\! \phi \left( 2  \varepsilon \gamma +\gamma \tilde{\Delta} \gamma  - \Sigma \gamma  + \gamma \tilde{\Sigma} \right)\mathrm{d}\Omega_j \Biggr]
\nonumber\\
= &- \sum\limits_{T_j \in \mathcal{T}} \int\limits_{T_j}\phi \Delta  ~\mathrm{d}\Omega_j,
\label{eq:WeakFormAfterPartialIntegration}
\end{align}
where $\mathbf{n}_j$ is the outer normal for $\partial \Omega_j$, the boundary of a given cell. For triangular cells each element boundary $\partial \Omega_j$ consists of three edges. Each edge, in turn, is either part of the geometric boundary $\partial \Omega$, or part of the collection of internal edges $\tau$. The geometric boundary $\partial \Omega$ can be further decomposed into the inflow boundary $\partial \Omega_-$ and the outflow boundary $\partial \Omega_+$, defined via
\begin{align}
   \partial \Omega_- &\equiv \left\{ {\RR} \in \partial \Omega ~|~\vF \cdot \mathbf{n}({\RR}) < 0 \right\} ,\\
    \partial \Omega_+ &\equiv \left\{ {\RR} \in \partial \Omega ~|~ \vF \cdot \mathbf{n}({\RR}) \geq 0 \right\}.
\end{align}
The three sets are sketched in Fig.~\ref{fig:mesh_example}(b). 
\begin{figure}[t]
    \centering
    \includegraphics{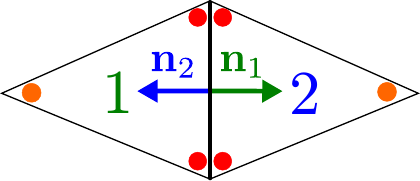}
    \caption{Sketch of two neighboring cells, 1 and 2, with the shared edge in between them. The two normal vectors $\mathbf{n}_1$ and $\mathbf{n}_2$ to the shared edge are chosen such that they are pointing out of their respective cell. The red dots indicate nodes that are involved in a surface integral over the edge, the orange dot is a node that does not contribute. The displacements of the dots from the actual nodal points of the cells are introduced only to illustrate the DG method.}
    \label{fig:GalerkinTwoCells}
\end{figure}
 In a discontinuous Galerkin method the unknown function $\gamma$ as well as the test function $\phi$ can have different values on the two sides of a given edge. Each internal edge $\tau_j \in \tau$ is shared by two cells and thus integrated over twice in the sum over all the boundary integrals. It can be shown \cite{Arnold2002} that summing this double integration over each edge allows to rewrite the first term on the left-hand side in Eq.~\eqref{eq:WeakFormAfterPartialIntegration} as
\begin{align}
&\sum\limits_{T_j \in \mathcal{T}} \int\limits_{ \partial\Omega_j}\!
\phi~(\gamma \vF)\cdot \mathbf{n}_j~\mathrm{d}s_j
\nonumber
=
\sum\limits_{\tau_j \in \tau }  \int\limits_{\tau_j}
 \left\{\gamma \vF \right\}\cdot  \left[\phi\right]~  \mathrm{d}\tau_j
 %\sum\limits_{\tau_j \in \tau} \int\limits_{\tau_j}
 %(\mathbf{n}_j \cdot \vF)~  \gamma  \phi  ~\mathrm{d}\tau_j.
 \nonumber
 \\
 &+
 \sum\limits_{s_j \in \partial \Omega_+} \int\limits_{s_j}
 (\mathbf{n}_j \cdot \vF)~  \gamma  \phi  ~\mathrm{d}s_j
 +\sum\limits_{ s_j\in \partial \Omega_-}  \int\limits_{s_j}
 (\mathbf{n}_j \cdot \vF)~  \gamma \phi  ~\mathrm{d}s_j,
 \label{eq:boundaryIntegral_Form}
\end{align}
where the summations on the right-hand side are over individual edges rather than closed boundaries of each triangle. 
The different brackets in the first term on the right-hand side denote the jump $[\dots]$ and average $\{\dots\}$ of a function over an edge shared by two cells 1 and 2. They are defined for vectors $\mathbf{a}$ and scalars $\phi$ as
\begin{align}
\left[ \mathbf{a} \right]
\equiv 
\mathbf{a}_1 \cdot \mathbf{n}_1 + \mathbf{a}_2 \cdot \mathbf{n}_2,
\quad
\left[ \phi \right]
\equiv 
\phi_1 \mathbf{n}_1 + \phi_2 \mathbf{n}_2,
\label{eq:JumpDefinition}
\\
\left\{ \mathbf{a} \right\}
\equiv 
\frac{1}{2} \left( \mathbf{a}_1 + \mathbf{a}_2 \right),
\quad
\left\{ \phi \right\}
\equiv 
\frac{1}{2} \left( \phi_1 + \phi_2 \right),
\label{eq:AverageDefinition}
\end{align}
with the index denoting the value in the respective cell, and the outward-pointing normal vector $\mathbf{n}_i$ in each of the two cells, see Fig.~\ref{fig:GalerkinTwoCells}. We note that in the literature the jump operator is sometimes written including the normal vector, i.e., $[\mathbf{a}] = [\mathbf{a} \cdot \mathbf{n}]$.
For edges on either the inflow or outflow boundary the function only has values on one side so no average or jump of the function values appears in the other two terms on the right-hand side of Eq.~\eqref{eq:boundaryIntegral_Form}.

Up until this point the derivation is identical for the retarded and advanced functions, $\gamma^\mathrm{R}$ and $\gamma^\mathrm{A}$, but start to be different in the following. For the retarded function, $\gamma^R$, a boundary condition $\gamma_\mathrm{B}$ has to be provided on the inflow boundary $\partial \Omega_-$. We address how these starting values are obtained in Sect.\ref{sec:BoundaryValuesSection}. 
On the outflow boundary $\partial \Omega_+$ and on internal edges $\tau$, the values of $\gamma^R$ are unknown and determined in the solution step. This corresponds to solving Eq.~\eqref{eq:GammaEquation} for the retarded function in the direction of $\vF$ with a boundary value at the start point of a given trajectory. For the advanced function $\gamma^\mathrm{A}$, $\partial \Omega_-$ and $\partial \Omega_+$ are swapped, corresponding to the stable integration direction opposite to $\vF$ in the stepping method. The remaining functions $\tilde{\gamma}^{R,A}$ are similarly swapped with respect to their non-tilde counterpart. The naming of $\partial \Omega_-$ ($\partial \Omega_+$) as inflow (outflow) boundary is thus only descriptive for $\gamma^\mathrm{R}$ and $\tilde{\gamma}^\mathrm{A}$ but follows the naming conventions established in literature.

Lastly, we note that for the imaginary-frequency Matsubara coherence function $\gamma^\mathrm{M}$, the propagation directions and flow boundaries correspond to the one for $\gamma^R$ ($\gamma^\mathrm{A}$) for positive (negative) Matsubara poles.

As discussed in literature \cite{Brezzi2004Dec}, better convergence and stability can be achieved for this weak form by replacing the average operator in the first term in Eq.~\eqref{eq:boundaryIntegral_Form} by the so-called \textit{upwind} value, denoted by a subscript $u$ and given by
\begin{align}
\left\{\gamma^R \vF \right\}_u 
\equiv
\left\{
	\begin{array}{ll}
		\gamma_1^R \vF   & \mbox{if } \vF \cdot \mathbf{n}_1 > 0 \\
		\gamma_2^R \vF & \mbox{if } \vF \cdot \mathbf{n}_1 < 0\\
		\left\{\gamma^R \right\} \vF & \mbox{if } \vF \cdot \mathbf{n}_1 = 0
	\end{array}
\right. .
\label{eq:UpwindDefinition_main}
\end{align}
The upwind value enforces a propagation of the function values across cell edges in the transport direction $\vF$, corresponding to the propagation along trajectories in the stepping method. 
A corresponding \textit{downwind} value with swapped signs in the inequalities in Eq.~\eqref{eq:UpwindDefinition_main} has to be used for the advanced function $\gamma^A$. This is similar to the exchange of $\partial\Omega_+$ and $\partial\Omega_-$ for the boundary values.   
The final weak form for the retarded function $\gamma^\mathrm{R}$ thus becomes
\begin{widetext}
\begin{align}
&i \hbar \! \sum\limits_{\tau_j \in \tau }  \int\limits_{\tau_j}
  \left\{\gamma^\mathrm{R} \vF \right\}_u \cdot \left[\phi\right]  ~\mathrm{d}\tau_j
 + i \hbar \! \sum\limits_{s_j \in \partial \Omega_+}  \int\limits_{s_j}
 (\gamma^\mathrm{R} \mathbf{n}_j \cdot \vF ) \phi  ~\mathrm{d}s_j
 - i \hbar\! \sum\limits_{T_j \in \mathcal{T}}  \int\limits_{T_j}
 ~(\gamma^\mathrm{R} \vF ) \cdot \nabla \phi~   ~\mathrm{d}\Omega_j
 \nonumber
  \\
&+ \sum\limits_{T_j \in \mathcal{T}}  \int\limits_{T_j} \left( 2  \varepsilon \gamma^\mathrm{R} + \gamma^\mathrm{R} \tilde{\Delta} \gamma^\mathrm{R} - \Sigma \gamma^\mathrm{R}  + \gamma^\mathrm{R} \tilde{\Sigma} \right) \phi~\mathrm{d}\Omega_j
= - \sum\limits_{T_j \in \mathcal{T}}  \int\limits_{T_j} \phi\Delta  ~\mathrm{d}\Omega_j
- i \hbar \sum\limits_{s_j \in \partial \Omega_-}  \int\limits_{s_j}(\mathbf{n}_j \cdot \vF )\phi~ \gamma^\mathrm{R}_B  ~\mathrm{d}s_j.
\label{eq:FullWeakForm_prelim}
\end{align}
\end{widetext}

\noindent 
Eq.~\eqref{eq:FullWeakForm_prelim} is of the form 
\begin{equation}
L(\gamma, \phi) = f(\phi),
\end{equation} 
where $f$ does not depend on the unknown function $\gamma$ while $L$ does. In the FEM language, $L$ is the bilinear form and $f$ the linear form of the weak formulation of Eq.~\eqref{eq:GammaEquation}. 
%No general strategy exist to find an exact \textit{weak solution} 
An approximate weak solution $\gamma_w^R$ to Eq.~\eqref{eq:FullWeakForm_prelim} can be found using a discontinuous Galerkin method. 
Such a method assumes that the approximate weak solution $\gamma_w^\mathrm{R}$ can, within each cell, be written as a sum of polynomials of finite order $k$.
Depending on $k$, there are $N_j$ independent degrees of freedom, or nodes, for a given cell $T_j$. In two dimensions, the common choice of $k=1$ (linear functions) or $k=2$ (quadratic functions) requires three and six nodes, respectively. For each of the $N_j$ nodes, there is an associated polynomial ${\phi}_i$. \textcolor{black}{In this paper we use simple Lagrange polynomials that satisfy}
\begin{align}
{\phi}_i(\mathbf{x}_j) = \delta_{ij}.
\end{align}
\textcolor{black}{Using this cell-wise basis} we can write
\begin{align}
\left. \gamma_w^\mathrm{R}\right|_{T_j}(\mathbf{R}) \approx \sum\limits_{i=1}^{N_j} a_{j,i} {\phi}_i(\mathbf{R}),
\label{eq:GalerkinMethodFunctionExpression}
\end{align}
where $a_{j,i}$ are expansion coefficients or weights. The function values on the nodes then fully determine $a_{j,i}$ and by Eq.~\eqref{eq:GalerkinMethodFunctionExpression} the function everywhere in the individual cell. The main difference to a continuous Galerkin method is that the values of $a_{j,i}$ on nodes shared between neighboring cells are independent, the global function can thus be discontinuous across cell edges. Such a discontinuous Galerkin method has been found to give better convergence for the transport equations and allows for higher mesh adaptability, see also the discussion in Ref.~\onlinecite{osti_4237082, Cockburn2000, Johnson2009Jan}. 

\textcolor{black}{We expand the self-energies in the same basis with weights $\Delta_{j,i}$ and $\Sigma_{j,i}$, while the test function $\phi$ used in Eq.~\eqref{eq:FullWeakForm_prelim} can be chosen to have the same form but with unit weights, $\phi = \sum_{i} {\phi_i}(\mathbf{x})$.}
% \begin{align}
% \left. \Delta_w^\mathrm{R}\right|_{T_j}(x) &\approx \sum\limits_{i=1}^{N_j} \Delta_{j,i} \hat{\phi}_i(x)
% \\
% \left. \Sigma_w^\mathrm{R}\right|_{T_j}(x) &\approx \sum\limits_{i=1}^{N_j} \Sigma_{j,i} \hat{\phi}_i(x),
% \label{eq:GalerkinMethodFunctionExpression}
% \end{align}

%\textcolor{black}{REference to some book here maybe.}

%In a continuous Galerkin method, the weights are by assumption equal on nodes shared between touching cells. 

Inserting Eq.~\eqref{eq:GalerkinMethodFunctionExpression} and the basis functions into Eq.~\eqref{eq:FullWeakForm_prelim} gives an equation system for the unknown coefficients $a_{j,i}$ for each cell. \textcolor{black}{Renumerating $a_{j,i} \rightarrow a_{i'}$ and combining all the local equations, we get a collection of scalar equations of the form} 
\begin{align}
\sum\limits_{j,k} a_{j} Q_{ijk} a_k + \sum\limits_{k} P_{ik} a_{k} = f_i, 
%\mathrm{diag}(\mathbf{a}) Q \mathbf{a} + P \mathbf{a} = \mathbf{f}.
\label{eq:WeakForm_MatrixEquation}
\end{align}
\textcolor{black}{where $i = 1,\dots, N$ is an index over the all $N$ nodes in our domain. In two dimensions, a triangulation with $N_T$ triangles and $N_j$ nodes per cell will have a total number of $N = N_T N_j$ nodes.}  
Here, the tensor $Q_{ijk}$ is due to the quadratic term $\gamma \tilde{\Delta} \gamma$ in Eq.~\eqref{eq:FullWeakForm_prelim}, while the remaining linear parts of $L(\gamma,\phi)$ are combined into a matrix $P_{ik}$. The nonlinearity prevents a solution of Eq.~\eqref{eq:WeakForm_MatrixEquation} by matrix inversion. Instead, the vector $\mathbf{a}$ is determined as an (approximate) zero of the residual vector $\mathbf{r}$, with components
\begin{align}
r_i(\mathbf{a}) \equiv \sum\limits_{j,k} a_{j} Q_{ijk} a_k + \sum\limits_{k} P_{ik} a_{k} - f_i, 
%R(\mathbf{a}, \phi) \equiv \mathrm{diag}(\mathbf{a})  Q \mathbf{a} + P \mathbf{a} - \mathbf{f}%L(\mathbf{a}, \phi) - f(\phi),
\end{align}
where again $i=1,\dots, N$ as in Eq.~\eqref{eq:WeakForm_MatrixEquation}.
\textcolor{black}{This approximate zero is found by iterative minimization}. We find that a numerically efficient strategy is Newton iteration using the Jacobian matrix $J_{ij}(\mathbf{a}, \phi)$ of Eq.~\eqref{eq:FullWeakForm_prelim}, which can be calculated from the residual via the Gateaux derivate
\begin{equation}
J_{ij}(\mathbf{a}, \phi, \delta a_j) \equiv 
%dR(\mathbf{a}, \phi) = 
\lim\limits_{h \rightarrow 0} \frac{ r_i(a_j + h~\delta a_j, \phi) - r_i(a_j,\phi) }{h}.
\end{equation}
Starting from a guess $\mathbf{a}_0$, one iterates the solution vector via Newton steps $\mathbf{a}_{n+1} = \mathbf{a}_n + (\delta \mathbf{a})_n$, where $(\delta \mathbf{a})_n$ is the solution to a linear equation system obtained by combining all $J_{ij}(\mathbf{a}_n, \phi) (\delta a_j)_n = -r_i(\mathbf{a}_n, \phi)$ for $i=1, \dots, N$.

\subsection{Boundary values}
\label{sec:BoundaryValuesSection}
As outlined in Sect.~\ref{sec:FiniteElementSection}, the weak form in Eq.~\eqref{eq:FullWeakForm_prelim} contains boundary values $\gamma_\mathrm{B}$, in the case of the retarded function on the inflow boundary $\partial \Omega_-$. Note that the inflow boundary will be different for different orientations of $\vF$. These boundary values correspond to the starting value of the coherence function at the start point of a trajectory in a stepping method.

We treat the entire boundary $\partial \Omega$, including the inflow boundary, as a collection of planar interfaces of finite length.  Scattering at such atomically sharp interfaces cannot be described within quasiclassical theory itself, the theory has to be supplemented by external boundary conditions. Using scattering theory, such conditions have been derived for specularly scattering interfaces \cite{eschrig_distribution_2000,Eschrig2009Oct,Zaitsev:1984,Millis:1988,Fogelstrom:2000,Zhao2004Oct} where the momentum parallel to the surface is conserved in the scattering process. The boundary conditions in the spin scalar case is at a given interface are expressed in terms of the normal-state scattering matrix
\begin{align}
\mathcal{S} = 
\begin{pmatrix}
\sqrt{R} & i\sqrt{D}
\\
i\sqrt{D} & \sqrt{R},
\end{pmatrix}
\end{align}
where $R$ and $D$ are, respectively, the reflectivity and transmittivity of the interface satisfying $R + D = 1$. 
The boundary condition for $\gamma^\mathrm{R}_\mathrm{B}$ is then of the form
\begin{align}
\gamma^\mathrm{R}_\mathrm{B} = \Gamma_\mathrm{out}(\gamma^\mathrm{R}_\mathrm{in,1}, \gamma^\mathrm{R}_\mathrm{in,2}, \mathcal{S} ), 
\label{eq:boundaryCondition}
\end{align}
where $\gamma^\mathrm{R}_\mathrm{in,i=1,2}$ are the incoming function inside and outside the system. Within the FEM nomenclature such boundary conditions are referred to as Dirichlet boundary conditions. In a discontinuous Galerkin method, they are weakly enforced through the last term on the right-hand side in Eq.~\eqref{eq:FullWeakForm_prelim}. On a given segment the entire function is determined by the values on the nodes, hence it is sufficient to apply the boundary condition at the nodes only.

At a fully reflective segment, $R=1$, specular scattering relates the outgoing function for an angle $\varphi_\mathrm{F}$ to the function incoming from within the system for a different angle $\varphi_\mathrm{F}'$. On a boundary segment with an outward-pointing normal $\mathbf{n}$ that spans an angle $\beta_n$ to the $x$-axis, we find
\begin{align}
\varphi_\mathrm{F}' = \pi - \varphi_\mathrm{F} + 2 \beta_n.
\end{align}
This relation holds in systems with a two-dimensional Fermi surface where the orientation of $\vF$ can be parametrized by one scalar parameter only. In three dimensions an analogous relation for two scalar parameters has to be used. The weak form in Eq.~\eqref{eq:FullWeakForm_prelim}, however, is valid in two and three dimensions if one replaces cells and edges with appropriately chosen polyhedra and surface planes.

\begin{figure}[t]
    \centering
    \includegraphics[trim={0 0 0cm 2.8cm},clip, width=0.45\textwidth]{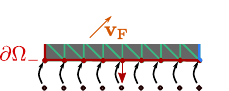}
    \caption{Coupling of each node (red dot) on the inflow boundary $\partial \Omega_-$ to virtual reservoir points (black empty circle).}
    \label{fig:reservoirs_sketch}
\end{figure}

Starting from a bulk guess as an incoming function, we use the incoming functions at an iteration $n$ to obtain the outgoing function in the next, $n+1$, iteration.
At a fully transparent segment, $D = 1$, the outgoing function solely depends an incoming function from a reservoir that is not directly simulated. This reservoir can for example be a superconducting contact, or a normal-metal reservoir. We model both cases by associating a virtual "reservoir" point with each node on the given boundary segment, see Fig.~\ref{fig:reservoirs_sketch}. 

We model a superconducting contact by using a bulk coherence function $\gamma_\mathrm{bulk}$ as the incoming coherence function from the reservoir. On each virtual point, the order parameter and superfluid momentum $\mathbf{p}_\mathrm{S}$ are then iterated such that we enforce a certain current density across the respective boundary segment. In the case of a normal-metal reservoir, a voltage bias (temperature bias) can be modelled by fixed electrochemical potential (temperature) of the reservoir so no update is required. This induces a nonequilibrium occupation at the system edge that propagates in the system.

\onecolumngrid

\begin{figure*}[t]
    \centering
    \includegraphics[width=\textwidth]{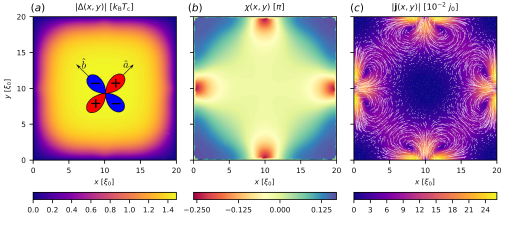}
    %\vspace{-20px}
    \caption{Phase crystal for $\Gamma = 5\cdot10^{-4} \pi \kb \tc$, unitary limit, and $T=0.12 \tc$. (a) $|\Delta|(x,y)$, the absolute value of the order parameter. (b) $\chi (x,y)$, the superconducting phase. In (c) the colormap shows the absolute value of the current $|\mathbf{j}(x,y)|$, while the flow direction is indicated by the superimposed vector field (white arrows).}
    \label{fig:phaseCrystals_example}
\end{figure*}

\twocolumngrid

For further details, we refer to our recent publications on nonequilibrium setups in (quasi) one-dimensional  models \cite{Seja2021Sep, Seja2022Mar}.
For intermediate values of the transparency, $D \in (0, 1)$, the boundary condition Eq.~\eqref{eq:boundaryCondition} gives a weighted combination of the two scenarios described above.

\subsection{The non-equilibrium distribution $\xx$}
A weak form for the non-equilibrium distribution $\xx$ can be obtained in a largely identical procedure as outlined for the coherence amplitude $\gamma$ in Sect.~\ref{sec:FiniteElementSection}. After partial integration over the term containing  the derivative $\vF \cdot \nabla x$ of splitting of the resulting boundary integrals has to be performed. The specification of boundary values and flow direction for $\xx$ ($\tilde{\xx}$) then follows that of $\gamma^\mathrm{R}$ (${\gamma}^\mathrm{A}$). The main difference to the coherence amplitude is that Eq.~\eqref{eq:XEquation} is clearly linear in $\xx$. Hence the solution of the resulting matrix equation can be obtained through matrix inversion rather than the iterative solution of a nonlinear problem.

\subsection{Selfconsistency \& Numerical aspects}
A self-consistent solution of the Eilenberger equation requires an iterative procedure of solving the underlying transport equation followed by an update of all self-energies. In equilibrium, this leads to the following recipe for the retarded functions:
\begin{enumerate}
    \item Provide a current guess of all self-energies in $\hat{h}^\mathrm{R}$, see Eq.~\eqref{eq:selfenergy_elements}, and boundary values $\gamma_\mathrm{B}^\mathrm{R}$ and $\tilde\gamma_\mathrm{B}^\mathrm{R}$.
    \item Find weak solutions $\gamma^\mathrm{R}_\mathrm{w}$ to Eq.~\eqref{eq:FullWeakForm_prelim}, and similarly $\tilde{\gamma}^\mathrm{R}_\mathrm{w}$, for all energies $\varepsilon$ and momentum angles $\varphi_\mathrm{F}$.
    \item Update all self-energies through Eqs.~\eqref{eq:gapequation} and \eqref{eq:ImpuritiesBorn} or \eqref{eq:ImpuritiesUnitary}.
    \item Update all boundary values $\gamma_\mathrm{B}^\mathrm{R}$ and $\tilde\gamma_\mathrm{B}^\mathrm{R}$ as described in Section~\ref{sec:BoundaryValuesSection}
    %\item Unless the self-energy updates are below a desired accuracy, return to step 1 using the self-energies obtained via step 3. 
\end{enumerate}
Non-equilibrium situations also require weak solutions $\xx_w$ and $\tilde{\xx}_w$ and updates of the Keldysh self-energies $\hat{h}^\mathrm{K}$ . The above procedure has to be repeated until a self-consistent solution is found, signalized for example by convergence of the self-energies up to a desired accuracy. Only then is charge-current conservation
\begin{align}
\nabla \cdot \mathbf{j}(\RR) = 0,
\label{eq:ChargeCurrent_conservation}
\end{align}
guaranteed everywhere in the system. For a given guess of all self-energies, the solution of Eq.~\eqref{eq:FullWeakForm_prelim} is an independent problem for each energy $\varepsilon$ and momentum orientation $\varphi_\mathrm{F}$. This makes the problem highly parallelizable.
%, for example via graphics cards, as in Ref.~{SuperConga}, or via MPI.
The solutions for all energies $\varepsilon$ and angles $\varphi_\mathrm{F}$ only need to be combined to obtain a new guess of the self-energies. For the FEM solution step we use the open-source package \textsc{Gridap} \cite{Badia2020}, while the meshes are created in \textsc{gmsh} \cite{Geuzaine2009Sep}. \textcolor{black}{Detail on the specific meshes used for the results presented in this paper can be found in a supplementary material online.}

\section{Results}
\label{sec:results}

To illustrate the strategies and the advantages of solving the Riccati equations with a FEM together with the self-consistency equations for the order parameter and scalar impurity self-energies, we present two examples. Both involve two-dimensional $d$-wave superconductors modeling for instance a single superconducting plane of a high-temperature superconductor. The first example is a closed system, a square island (see Fig.~\ref{fig:phaseCrystals_example}), where the superconducting crystal axes are rotated 45$^\circ$ relative to the main axes of the square. For this orientation of the order parameter relative to the edges, below a phase transition temperature $T^* \approx 0.17 \tc$, time-reversal symmetry has been predicted to be broken in an unusual way \cite{Hakansson2015Sep,Holmvall2017,Holmvall2018Jun,Holmvall2020Jan,Wennerdal2020Nov,Chakraborty2022Apr}. The characteristic of this state, referred to as a phase crystal \cite{Holmvall2020Jan}, is a non-trivial, structured ordering of the superconducting phase $\chi(\RR)$ resulting in patterns of current flow consisting of loops near the edges. In the present example we show the correctness of the FEM solution by studying the detrimental influence of scalar impurity scattering within a self-consistent homogeneous scattering model. In the second example we study an open system, a current biased $d$-wave superconducting bridge with an inhomogeneous density of impurities, modeling experiments on grooved Dayem bridges \cite{Trabaldo2020Mar,Trabaldo2020Jun}. In this case we also demonstrate how to self-consistently enforce a current boundary condition at the source and drain leads and current conservation across the bridge. This nicely illustrates the effect of current focusing and we predict the spatial variation of the superconducting phase over the bridge. \textcolor{black}{In both example setups we assume that scattering at non-transparent boundaries is specular, as discussed in Sect.~\ref{sec:BoundaryValuesSection}.}

\begin{figure}[t]
    \centering
    \includegraphics{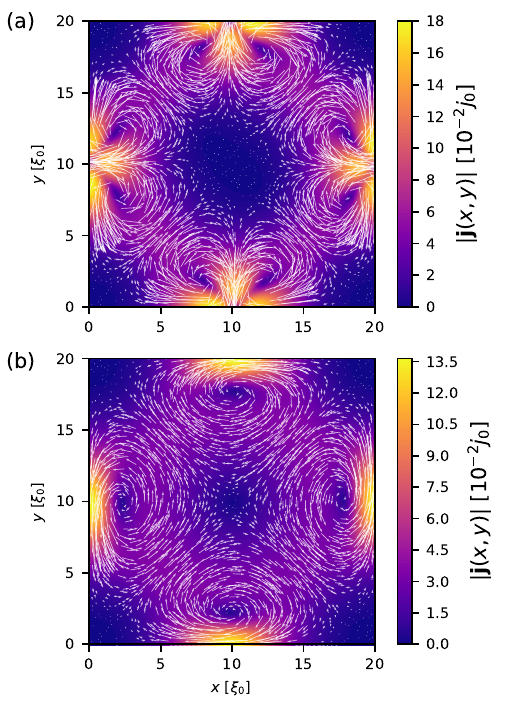}
    \vspace{-20px}
    \caption{Absolute value of the current flow $|\mathbf{j}(x,y)|$ (colormap), with the flow direction indicated by the superimposed vector field (white arrows). Note that the arrows are only illustrative and not scaled to to the magnitude of the flow. The impurities scatter in the unitary limit with a pair breaking parameter $\Gamma = 4.55\cdot10^{-2}\pi \kb \tc$ (mean free path $\ell\approx 22\xi_0$). In (a), $T=0.03 T_\mathrm{c}$ and in (b) $T = 0.07 T_\mathrm{c}$, corresponding to the second and right-most brown diamond in Fig.~\ref{fig:phaseCrystals_Gamma}(b).}
    \label{fig:phaseCrystals_flow_example}
\end{figure}

\subsection{Effect of scalar impurities on phase crystals}
Consider the square $d$-wave superconducting island in Fig.~\ref{fig:phaseCrystals_example}. When the crystal axes are misaligned by 45$^\circ$ relative to the surface norms, the $d$-wave order parameter is suppressed to zero at the edges, see Fig.~\ref{fig:phaseCrystals_example}(a). This reflects the formation of zero-energy surface Andreev bound states \cite{Hu1994,kashiwaya_tunnelling_2000, lofwander_andreev_2001}. At temperatures below $T^* \approx 0.17 \tc$ a second-order phase transition  was recently predicted, where time-reversal symmetry is broken and a non-uniform state appears with circulating currents near the edges \cite{Hakansson2015Sep}. The influence of mesoscopic roughness of the boundaries \cite{Holmvall2017} and magnetic field \cite{Holmvall2018Jun} have been investigated earlier. Within a tight-binding model, also the influence of band structure was investigated \cite{Wennerdal2020Nov}. Recently, an investigation of strong correlations within a Gutzwiller approximation preventing double occupancy on each tight-binding site was reported \cite{Chakraborty2022Apr}. In the latter study it was shown that Anderson disorder suppresses $T^*$, but surprisingly that strong correlations weakens the influnce of this disorder. In our example, we study the suppression of $T^*$ within a different model for the scalar impurities. In our quasiclassical treatment we self-consistently compute the impurity self-energy in a homogeneous scattering model for two types of disorder: 
the Born limiting of weak scattering, Eq.~\eqref{eq:ImpuritiesBorn}, and the unitary limit of diverging scattering potential strength, Eq.~\eqref{eq:ImpuritiesUnitary}. 

\begin{figure}[t]
    \centering
    \includegraphics{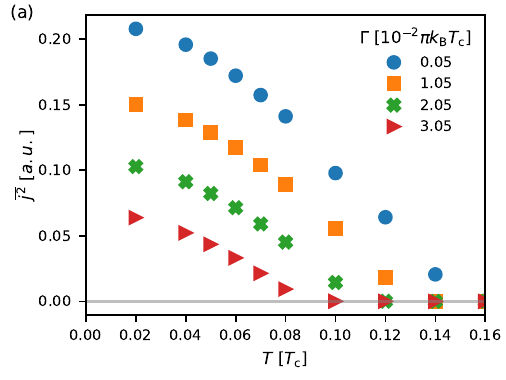}
    \vspace{-10px}
    \includegraphics{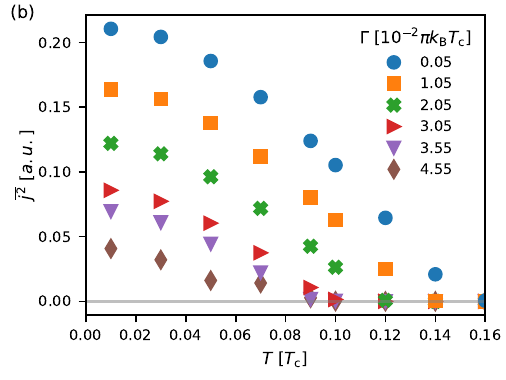}    
    \caption{Area-integrated current flow, see Eq.~\eqref{eq:integratedCurrent}, at varying temperature for different impurity concentrations in the Born limit (upper figure) and in the unitary scattering limit (lower figure).
    %\textcolor{black}{More data points on the way and improved accuracy in clean case needed.}
    }
    \label{fig:phaseCrystals_Gamma}
\end{figure}

In both cases, larger impurity concentrations suppress the current flow up to a critical value $\Gamma^*$ at which the phase crystal is no longer energetically favorable and a uniform superconducting state is formed instead. 
Examples of current flow patterns for two temperatures, $T=0.03 \tc$ and $T=0.07\tc$, in the unitary limit for a larger value of the impurity concentration is shown in Fig.~\ref{fig:phaseCrystals_flow_example}. As compared to the almost clean case in Fig.~\ref{fig:phaseCrystals_example}(c), for such impurity concentrations, such as Fig.~\ref{fig:phaseCrystals_flow_example} (normal state mean free path $\ell\approx 22 \xi_0$), the pattern remains largely unchanged at the lower temperature while the absolute value of the current is reduced. 
At the higher temperature for the same impurity concentration, a pattern with a reduced number of loops is observed, see Fig.~\ref{fig:phaseCrystals_flow_example}(b).

\begin{figure}[t]
    \centering
    \includegraphics[width=\columnwidth]{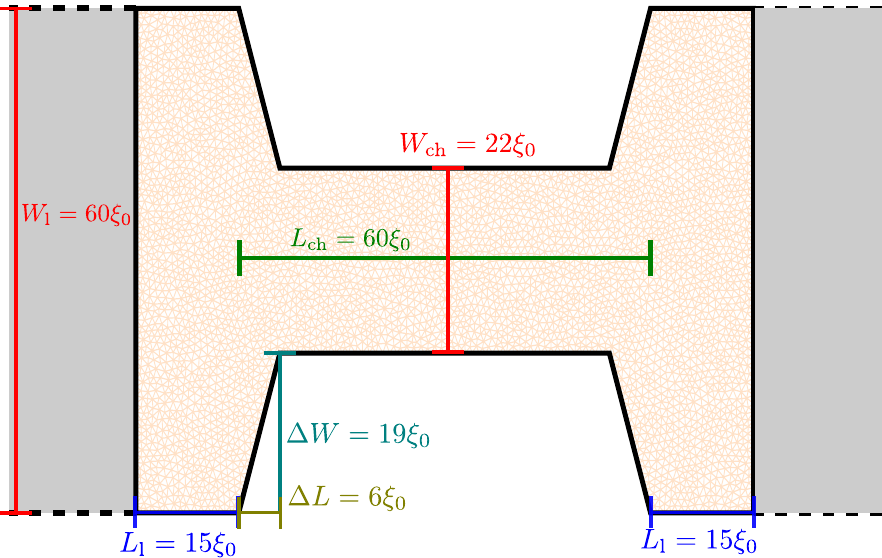}
    \caption{Sketch of an approximate Dayem-bridge geometry. A wide, transparent contact of width $W_l = 60 \xi_0$ connects, via a lead of length $L_\mathrm{l} = 15\xi_0$, to a narrow channel of width $W_\mathrm{ch} = 22\xi_0$ and length $L_\mathrm{ch} = 60 \xi_0$. At the connection of lead and channel, the width of the system linearly changes by $\Delta W = 19 \xi_0$ over a length of $\Delta L = 6 \xi_0$. Only the orange-shaded area is simulated, the grey areas to the left and right indicate the superconducting reservoirs, connected here with a fully transparent interface ($D=1$).}
    \label{fig:dayemSketch}
\end{figure}

As a measure of the total amount of current flow in the system we use the integrated quantity
\begin{align}
\overline{j^2} \equiv \int\limits_\Omega  |\mathbf{j}(x,y)|^2~\mathrm{d} \Omega.
\label{eq:integratedCurrent}
\end{align}
This average measure depends on temperature, impurity concentration, and type of impurities (Born or unitary), as summarized in Fig.~\ref{fig:phaseCrystals_Gamma}. For the relatively clean case (blue circles), the suppression of $\overline{j^2}$ with temperature leads to an estimate of the transition temperature $T^*$. For $T<T^*$ the phase crystal is energetically favorable \cite{Hakansson2015Sep}.
The corresponding temperature dependencies of $\overline{j^2}$ for increasing impurity concentration $\Gamma$ is shown in Fig.~\ref{fig:phaseCrystals_Gamma}(a) for the Born limit and Fig.~\ref{fig:phaseCrystals_Gamma}(b) for the unitary limit. Generally, for the same temperature and impurity concentration the currents are smaller for scattering in the Born limit.
%The critical value $\Gamma^*(T)$ is also consistently lower this type of impurity scattering.
It is well-known that the surface Andreev bound states are less broadened in the unitary limit than in the Born limit \cite{Poenicke1999}. As a result of the larger broadening of the Andreev bound states in the Born limit the spontaneous currents are energetically less favourable.
%Indeed, we find that a comparison of the free energy shows ...
For a temperature $T=0.1\tc$ the currents are well developed in the clean case, with a typical wavelength for the current loops of $\sim 12\xi_0$ (two loops with opposite circulations \cite{Hakansson2015Sep}). With increasing $\Gamma$, first the pattern changes to a single loop and then vanishes at this temperature at a mean free path of order $\ell\sim 30\xi_0$ in both the unitary and Born limits.
\textcolor{black}{
These results indicate that the phase diagram in $\Gamma-T$-space can be rich and should be investigated in further detail, but this is beyond the scope of the present paper.}

\subsection{Focussing of current flow in a Dayem bridge}

\begin{figure}[t]
    \centering
    \includegraphics{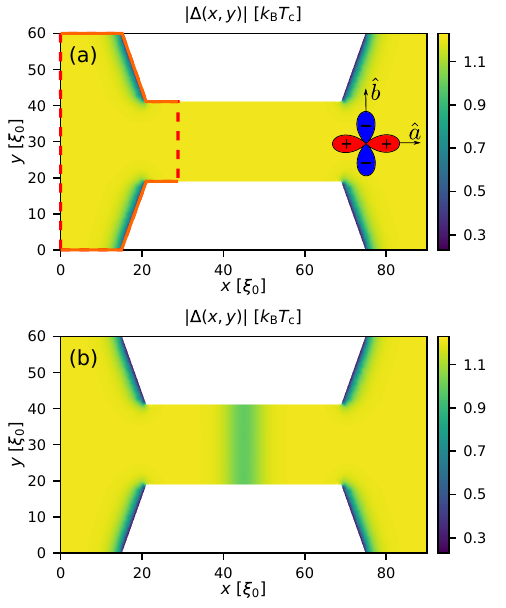}
    \caption{Reduction of the order parameter by the reduced mean free path in the central groove of the channel. (a) Constant impurity concentration $\Gamma = \Gamma_0$, (b) Spatial variation according to Eq.~\eqref{eq:ImpurityGroove}. In (a), an example for a contour described in the text and used for Eq.~\eqref{eq:currentConservationCheck} is superimposed. The integrated current incoming at the left edge and through the dashed line parallel to the $y$ axis on the right edge has to be conserved.}
    \label{fig:Delta_noFlow}
\end{figure}

\begin{figure*}[t]
    \centering
    \includegraphics[width=\textwidth]{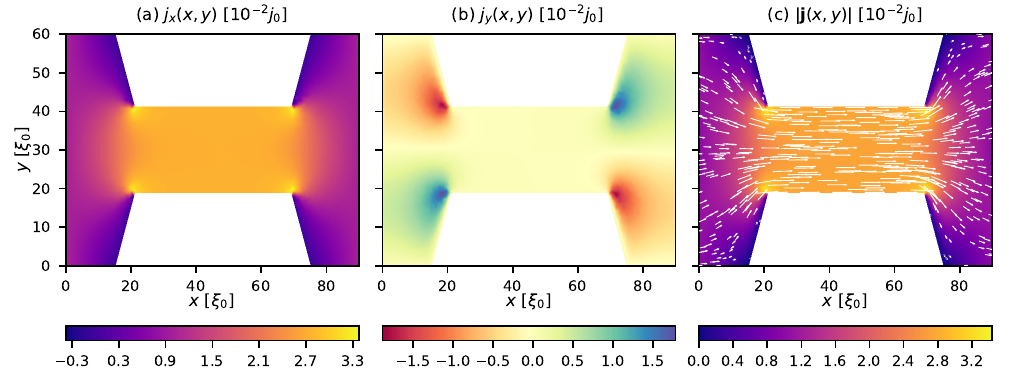}
    \caption{Current flow with weak link for a uniform boundary current density of $\mathbf{j} = 0.01 j_0 \hat{x}$. (a) The $x$ component $j_x$, (b) the $y$ component $j_y$, (c) absolute value $|\mathbf{j}|$ with the vector field $\mathbf{j}$ superimposed.}
    \label{fig:currentFlow_weakLink}
\end{figure*}

As an example of an open system with current flow we consider a $d$-wave superconducting bridge, inspired by recent experimental realizations of Dayem bridges \cite{Trabaldo2020Mar,Trabaldo2020Jun}. Fig.~\ref{fig:dayemSketch} shows the model for such a bridge. It consists of two superconducting reservoirs smoothly connected to a narrow channel of reduced width. The $d$-wave order parameter is in this case aligned with the bridge, such that the lobes are along $x$- and $y$-directions. Additionally, a groove can be etched into the channel. %creating a weak link at low temperatures. 
To model this groove we include a position-dependent impurity concentration of Gaussian shape in the center of the channel, 
\begin{align}
\Gamma(x) = \Gamma_0 + \Gamma_\mathrm{peak}  e^{-(x-x_\mathrm{center})^2/2 w_\mathrm{Groove}^2}.
\label{eq:ImpurityGroove}
\end{align}
In the results presented here we use $x_\mathrm{center} = 45\xi_0$, at the center of the channel, and $w_\mathrm{Groove} = \sqrt{0.2} \xi_0$. With these parameters the impurity concentration has a smooth Gaussian profile over a characteristic scale of roughly $10 \xi_0$ around $x_0$. For impurity concentrations, we choose $\Gamma_0 = 5\cdot10^{-4} \pi \kb \tc$ and $\Gamma_\mathrm{peak} = 0.05 \pi \kb \tc$, corresponding to a normal-state mean free path of $\ell_0 = 2000 \xi_0$ and $\ell_\mathrm{peak} = 20 \xi_0$, respectively. Unless noted otherwise, all results below are for a temperature of $T = 0.7 \tc$.

The reduced mean free path in the groove results in a suppression of the order parameter by around twenty percent in the dirty region in the center of the channel already in equilibrium, as compared to the case without an impurity concentration profile, see in Fig.~\ref{fig:Delta_noFlow}. Note also that the order parameter is, in both cases, partially suppressed at the tilted edges in the transition area between the wide contact and the narrow channel a result of the partially pair-breaking nature of the walls there.

We now enforce a constant boundary current density of $j_x = j_\mathrm{b} = 0.01 j_0$ at the left and right edges of the system. Upon a self-consistent calculation, the current that is flowing in the $x$ direction from the left reservoir gets spatially focused to flow into the narrow channel.

\begin{figure}[t]
    \centering
    \includegraphics{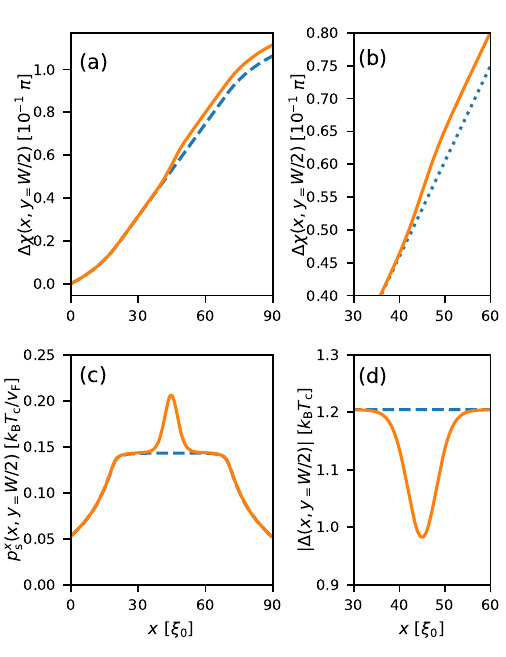}
    \vspace{-1cm}
    \caption{Variation of the superconducting order parameter along a line cut in the center of the channel, $y \approx W_l/2$, %, for the two scenarios described above.
    \textcolor{black}{with $\Gamma_\mathrm{peak} =0$ (dashed blue) and $\Gamma_\mathrm{peak} = 5\cdot10^{-2} \pi \kb \tc$ (solid orange).}
    }
    \label{fig:Delta_weakLink}
\end{figure}

\begin{figure}[t]
    \centering
    \includegraphics{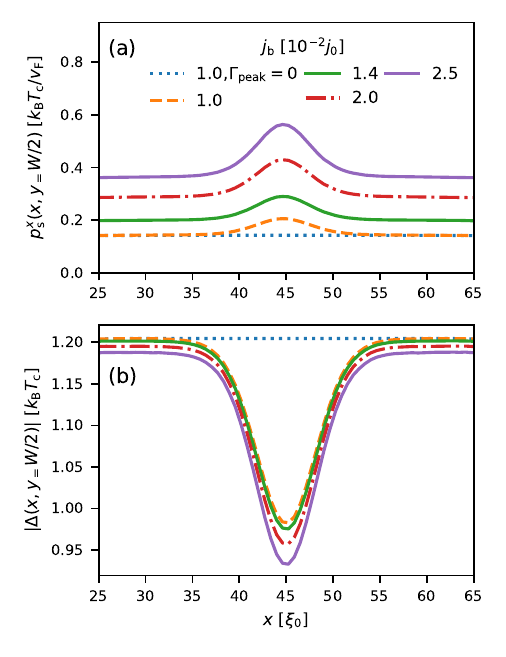}
    \caption{Variation of the superconducting order parameter along a line cut in the center of the channel, $y \approx W_l/2$ (see Fig. \ref{fig:dayemSketch}), \textcolor{black}{for $\Gamma_\mathrm{peak} =0$ (dotted blue) and $\Gamma_\mathrm{peak} = 5\cdot10^{-2} \pi \kb \tc$ (other lines) at different values of the boundary current $j_\mathrm{b}$}. (a) The $x$ component of the superflow $  \mathbf{p}_\mathrm{s}^x = (\hbar/2) \partial_x \chi(x,y) $, and (b) the suppression of the order parameter in the groove.}
    \label{fig:Delta_psComparison}
\end{figure}

By integration of Eq.~\eqref{eq:ChargeCurrent_conservation} and application of Gauss' theorem in two-dimensions we can reinterpret the condition of local current conservation as
\begin{equation}
\int_{\mathrm{int}~\mathcal{C}} \nabla \cdot \mathbf{j} ~ \mathrm{d}\Omega= \int_\mathcal{C} \mathbf{n} \cdot \mathbf{j}~\mathrm{d}s = 0,
\label{eq:currentConservationCheck}
\end{equation}
where $\mathcal{C}$ is a closed contour within our system and $\mathrm{int}~\mathcal{C}$ the enclosed area. An example for such a contour is shown in Fig.~\ref{fig:Delta_noFlow}(a). We always include the left system edge as well as the upper and lower edges of the geometry. The last part of the contour is then a line parallel to the $y$-axis at different distances to the left edge.  

We can then integrate $j_x(x,y)$, the $x$ component of the current, over the local height $W(x)$ of the structure,
\begin{equation}
I_x(x) \equiv \int_0^{W(x)} j_x(x,y)~ \mathrm{d} y.
\label{eq:current_yIntegrated}
\end{equation}
Since no current can flow out of the reflective walls, the integrated current flow through the left and right contour edge, indicated as dashed lines in Fig.~\ref{fig:Delta_noFlow}(a), has to be conserved throughout the system. In the results presented here, this is the case up to a relative error of $\delta I_x < 2\cdot 10^{-2} I_x (0)$.  

A self-consistent result of the spatially redistributed current flow is shown Fig.~\ref{fig:currentFlow_weakLink}. The focusing of the current from the wide lead into the narrow channel is clearly visible. For the same imposed boundary current the flow pattern is basically identical for both grooved and non-grooved channels since the $y$-integrated current, Eq.~\eqref{eq:current_yIntegrated}, is conserved across the weak link in both cases. However, the suppression of the order parameter in the grooved Dayem bridge leads to an increased phase gradient in the channel. This can be more clearly seen by plotting the phase $\chi$ and $x$ component of the superfluid momentum 
\begin{align}
p_\mathrm{s}^x \equiv \frac{\hbar}{2} \partial_x \chi(x,y),    
\end{align}
along a line cut in the $x$ direction for constant $y = W_\mathrm{l}/2$, in the middle of the channel. Both quantities are shown in Fig.~\ref{fig:Delta_weakLink} for a non-grooved (blue dashed lines) and a grooved Dayem bridge (orange solid lines). In order to carry the same current in the $x$ direction the phase gradient is enhanced in the groove region to compensate for the reduced order parameter. For increasing values of the boundary current, the suppression of the order parameter amplitude and the enhancement of the superfluid momentum become more and more pronounced, see Fig.~\ref{fig:Delta_psComparison}. Beyond the highest boundary current $j_b>0.025j_0$ superconductivity breaks down locally in the channel in the sense that the order parameter is locally suppressed to zero during the self-consistency iteration. After that we find no self-consistent stationary solution for the order parameter in the weak link.

%\input{Text/discussion}
%\cleardoublepage
\section{Discussion and Outlook}
\label{sec:discussion}
In this paper, we have presented a reformulation of Eilenberger's quasiclassical theory of superconductivity in terms of a discontinuous Galerkin method.  We applied the method, firstly, to study the influence of scalar impurity scattering on phase crystals in $d$-wave superconductors, and, secondly, to investigate the current flow through a geometric constriction in the form of a Dayem bridge. 

In the first case, we find that scalar impurity scattering suppresses the spontaneous flow patterns of phase crystals. As a result the average flow gets suppressed compared to a clean $d$-wave superconductor and the characteristic temperature $T^*$, where the phase crystal is destroyed, is reduced. For the same impurity concentration the spontaneous flow is lower for impurities in the Born limit compared to the limit of unitary scattering. We attribute this to the increased broadening of the surface Andreev bound states in the Born limit which makes spontaneous flow energetically less favourable. 

Our second example shows the strength of the FEM method in the application to open superconducting systems in the presence of current flow. Generally, the constriction in the Dayem bridge geometry leads to a increased phase gradient in the narrow channel. In the case of a grooved Dayem bridge, the reduced mean free path in the Groove requires a strong increase in the phase gradient within the groove region in order to carry the same current. As a result, the groove can lead to a quick suppression of superconductivity with increasing current although the flow itself is non-dissipative.

The method presented here gives a powerful and versatile numerical technique to study conventional and unconventional superconductors with arbitrary mean free path in two- and three-dimensional geometries. The adaptability of a FEM to arbitrary geometries allows the investigation of realistic device geometries and experimental setups. An extension to the case of full spin-structure of the quasiclassical propagators follows identical steps as discussed here and would allow to study for example the effects of spin-orbit interactions and other unconventional superconductors. \textcolor{black}{Another extension, relevant for, e.g., the situation in the Dayem bridge when we can not find self-consistent stationary solutions at high imposed currents, would be to include time-dependence.}

\acknowledgements

We thank J. Wiman and O. Shevtsov for early discussions on the possibility of applying the finite element method within quasiclassical theory, and acknowledge the Swedish research council for financial support. The computations were enabled by resources provided by the Swedish National Infrastructure for Computing (SNIC) at NSC partially funded by the Swedish Research Council through grant agreement no. 2018-05973.

\section*{Appendix: Comparison of numerical performance}

To benchmark the numerical performance of the DG method presented in this paper, we use a simple example in one dimension. Specifically, we assume a clean system with an order-parameter profile of the form
\begin{align}
    \Delta(x) = \Delta_0 \tanh \frac{x-x_\mathrm{mid}}{2\xi_0},
    \label{eq:DeltaProfile_example}
\end{align}
where $\Delta_0 = 1.5 \kb \tc$, $x_\mathrm{mid} = 7.5 \xi_0$, and let $x \in [0, 15\xi_0]$. This profile is similar to that close to a fully reflective, pair-breaking interface to a $d$-wave superconductor. For this scenario, we compare
\begin{enumerate}
    \item the ''stepping``method, see Sect.~\ref{sec:trajectory_method},
    \item an implicit midpoint method, a well-known finite-difference scheme, see also Appendix C in Ref.~\cite{Holmvall2022May},
    \item the DG method presented here, for different order $k$ of the approximating polynomials.
\end{enumerate}
In all cases, we assume a uniform grid of $N$ points (or nodes), which gives rise to a step size or cell size $h = L / N = 15 \xi_0 / N$. As a starting value, we assume $\gamma(x=0)= 0$ at the start of the trajectory.

\begin{figure}[t]
    \centering
    \includegraphics{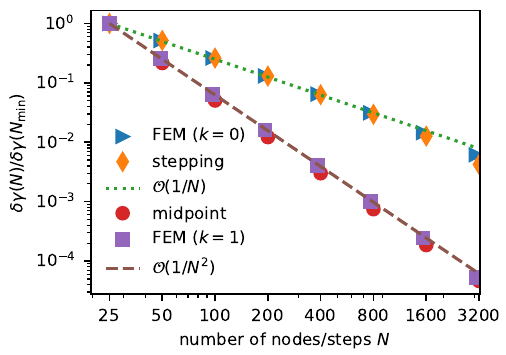}
    \caption{Comparison of $\delta \gamma(N)$, see Eq.~\eqref{eq:deltaGamma_definition}, as a function of the number of grid points $N$, for the different approaches. Addtionally, the expected error scaling for $1/N$ (dotted) and $1/N^2$ (dashed).}
    \label{fig:scaling_comparison}
\end{figure}

In the absence of an analytic solution for the order-parameter profile in Eq.~\eqref{eq:DeltaProfile_example}, we take the numerical solution of the stepping method for a very dense grid, using $N_\mathrm{max}=51200$ points, as the reference solution. As a measure of the numerical error, we use the summed difference
\begin{align}
    \delta \gamma(N) = \sum\limits_{i=1}^N~|\gamma_N(x_i) - \gamma_{N_\mathrm{max}}(x_i)|,
    \label{eq:deltaGamma_definition}
\end{align}
where we choose the $x_i$ such that they are included in the grid for all values of $N$. A plot of the error scaling as function of $N$, normalized to $\delta \gamma (N_\mathrm{min})$ the error for the grid with the lowest number of points $N_\mathrm{min} = 25$, is shown in Fig.~\ref{fig:scaling_comparison}. 
%\columnbreak

%\pagebreak

For our benchmark problem, we find that the error $\delta \gamma(N)$ is of order $\mathcal{O}(1/N)$ for both the stepping method and a corresponding DG method with piecewise constant polynomials $(k=0)$ as approximating functions. Once we use linear approximating functions $(k=1)$, we get an improved scaling behavior of order $\mathcal{O}(1/N^2)$, which we also find in the midpoint method. In both approaches, this improved error scaling is only found if the self-energy profile is also modeled as a linearly varying function in a given step length or cell. For a DG method with quadratic polynomials ($k=2$), we find an error scaling of order $\mathcal{O}(1/N^3)$ (not shown in Figure \ref{fig:scaling_comparison}). From the literature on DG methods we expect that for approximating polynomials up to order $k$, the error scaling is of order $\mathcal{O}((1/N)^{k+1})$ for linear hyperbolic problems on uniform grids \cite{Cockburn2000}. While some literature exists on DG methods for non-linear problems \cite{Hesthaven2008}, it appears that a more thorough analysis of the error scaling for the nonlinear Riccati equation is an open question.

The computational time needed to solve this example in one dimension scales linearly with the number of nodes $N$ and is thus of the same order of magnitude for all presented methods. In two dimensions, the midpoint method should scale as $N^2$ on a square geometry while the equivalent FEM is predicted to scale as $Nd^2$, where $d$ is the so-called band width of entries around the diagonal~\cite{Johnson2009Jan}. Since $d$ is determined by the upwind or downwind coupling of neighboring cells, $d$ can be made small compare to $N$ by numerical optimization.

%It is well-known that a Riccati equation is equivalent to a homogeneous, linear equation system of first-order ordinary differential equations \cite{Reid1972Jan}, or equivalently, that the coherence amplitude $\gamma$ can be constructed from solutions to the Andreev equations \cite{Shelankov2000Mar}. However, if this implies that the
% non-linear Riccati equation under consideration here, the 
%Obviously, these results are not guaranteed to directly transfer to our non-linear problem but to the best of our knowledge no explicit mathematical analysis has been done for the complex-valued Riccati equation. \cite{Cockburn2000}

%\cleardoublepage 
\bibliography{references}

\begin{thebibliography}{46}%
\makeatletter
\providecommand \@ifxundefined [1]{%
 \@ifx{#1\undefined}
}%
\providecommand \@ifnum [1]{%
 \ifnum #1\expandafter \@firstoftwo
 \else \expandafter \@secondoftwo
 \fi
}%
\providecommand \@ifx [1]{%
 \ifx #1\expandafter \@firstoftwo
 \else \expandafter \@secondoftwo
 \fi
}%
\providecommand \natexlab [1]{#1}%
\providecommand \enquote  [1]{``#1''}%
\providecommand \bibnamefont  [1]{#1}%
\providecommand \bibfnamefont [1]{#1}%
\providecommand \citenamefont [1]{#1}%
\providecommand \href@noop [0]{\@secondoftwo}%
\providecommand \href [0]{\begingroup \@sanitize@url \@href}%
\providecommand \@href[1]{\@@startlink{#1}\@@href}%
\providecommand \@@href[1]{\endgroup#1\@@endlink}%
\providecommand \@sanitize@url [0]{\catcode `\\12\catcode `\$12\catcode
  `\&12\catcode `\#12\catcode `\^12\catcode `\_12\catcode `\%12\relax}%
\providecommand \@@startlink[1]{}%
\providecommand \@@endlink[0]{}%
\providecommand \url  [0]{\begingroup\@sanitize@url \@url }%
\providecommand \@url [1]{\endgroup\@href {#1}{\urlprefix }}%
\providecommand \urlprefix  [0]{URL }%
\providecommand \Eprint [0]{\href }%
\providecommand \doibase [0]{https://doi.org/}%
\providecommand \selectlanguage [0]{\@gobble}%
\providecommand \bibinfo  [0]{\@secondoftwo}%
\providecommand \bibfield  [0]{\@secondoftwo}%
\providecommand \translation [1]{[#1]}%
\providecommand \BibitemOpen [0]{}%
\providecommand \bibitemStop [0]{}%
\providecommand \bibitemNoStop [0]{.\EOS\space}%
\providecommand \EOS [0]{\spacefactor3000\relax}%
\providecommand \BibitemShut  [1]{\csname bibitem#1\endcsname}%
\let\auto@bib@innerbib\@empty
%</preamble>
\bibitem [{\citenamefont
  {Eilenberger}(1968)}]{eilenberger_transformation_1968}%
  \BibitemOpen
  \bibfield  {author} {\bibinfo {author} {\bibfnamefont {G.}~\bibnamefont
  {Eilenberger}},\ }\bibfield  {title} {\bibinfo {title} {Transformation of
  {Gorkov}'s equation for type {II} superconductors into transport-like
  equations},\ }\href {https://doi.org/10.1007/BF01379803} {\bibfield
  {journal} {\bibinfo  {journal} {Z. Physik}\ }\textbf {\bibinfo {volume}
  {214}},\ \bibinfo {pages} {195} (\bibinfo {year} {1968})}\BibitemShut
  {NoStop}%
\bibitem [{\citenamefont {Larkin}\ and\ \citenamefont
  {Ovchinnikov}(1968)}]{Larkin1969}%
  \BibitemOpen
  \bibfield  {author} {\bibinfo {author} {\bibfnamefont {A.}~\bibnamefont
  {Larkin}}\ and\ \bibinfo {author} {\bibfnamefont {Y.~N.}\ \bibnamefont
  {Ovchinnikov}},\ }\bibfield  {title} {\bibinfo {title} {{Quasiclassical
  Method in the Theory of Superconductivity}},\ }\href
  {http://jetp.ras.ru/cgi-bin/e/index/e/28/6/p1200?a=list} {\bibfield
  {journal} {\bibinfo  {journal} {Zh. Eksp. Teor. Fiz.}\ }\textbf {\bibinfo
  {volume} {55}},\ \bibinfo {pages} {2262} (\bibinfo {year} {1968})},\ \bibinfo
  {note} {[Sov. Phys. JETP {\bf 28}, 1200, 1969]}\BibitemShut {NoStop}%
\bibitem [{\citenamefont {Eliashberg}(1971)}]{Eliashberg1971}%
  \BibitemOpen
  \bibfield  {author} {\bibinfo {author} {\bibfnamefont {G.}~\bibnamefont
  {Eliashberg}},\ }\bibfield  {title} {\bibinfo {title} {Inelastic electron
  collisions and nonequilibrium stationary states in superconductors},\ }\href
  {http://jetp.ras.ru/cgi-bin/e/index/e/34/3/p668?a=list} {\bibfield  {journal}
  {\bibinfo  {journal} {Zh. Eksp. Teor. Fiz.}\ }\textbf {\bibinfo {volume}
  {61}},\ \bibinfo {pages} {1254} (\bibinfo {year} {1971})},\ \bibinfo {note}
  {[Sov. Phys. JETP {\bf 34}, 668 (1972)]}\BibitemShut {NoStop}%
\bibitem [{\citenamefont {Reed}\ and\ \citenamefont
  {Hill}(1973)}]{osti_4491151}%
  \BibitemOpen
  \bibfield  {author} {\bibinfo {author} {\bibfnamefont {W.~H.}\ \bibnamefont
  {Reed}}\ and\ \bibinfo {author} {\bibfnamefont {T.~R.}\ \bibnamefont
  {Hill}},\ }\bibfield  {title} {\bibinfo {title} {Triangular mesh methods for
  the neutron transport equation},\ }in\ \href
  {https://www.osti.gov/biblio/4491151} {\emph {\bibinfo {booktitle}
  {Conference proceedings of the ''National topical meeting on mathematical
  models and computational techniques for analysis of nuclear systems``}}}\
  (\bibinfo {year} {1973})\BibitemShut {NoStop}%
\bibitem [{\citenamefont {Lasaint}\ and\ \citenamefont
  {Raviart}(1974)}]{Lasaint1974Jan}%
  \BibitemOpen
  \bibfield  {author} {\bibinfo {author} {\bibfnamefont {P.}~\bibnamefont
  {Lasaint}}\ and\ \bibinfo {author} {\bibfnamefont {P.~A.}\ \bibnamefont
  {Raviart}},\ }\bibfield  {title} {\bibinfo {title} {{On a Finite Element
  Method for Solving the Neutron Transport Equation}},\ }in\ \href
  {https://doi.org/10.1016/B978-0-12-208350-1.50008-X} {\emph {\bibinfo
  {booktitle} {{Mathematical Aspects of Finite Elements in Partial Differential
  Equations}}}}\ (\bibinfo  {publisher} {Academic Press},\ \bibinfo {address}
  {Cambridge, MA, USA},\ \bibinfo {year} {1974})\ pp.\ \bibinfo {pages}
  {89--123}\BibitemShut {NoStop}%
\bibitem [{\citenamefont {Kaper}\ \emph {et~al.}(1974)\citenamefont {Kaper},
  \citenamefont {Leaf},\ and\ \citenamefont {Lindeman}}]{osti_4237082}%
  \BibitemOpen
  \bibfield  {author} {\bibinfo {author} {\bibfnamefont {H.~G.}\ \bibnamefont
  {Kaper}}, \bibinfo {author} {\bibfnamefont {G.~K.}\ \bibnamefont {Leaf}},\
  and\ \bibinfo {author} {\bibfnamefont {A.~J.}\ \bibnamefont {Lindeman}},\
  }\bibfield  {title} {\bibinfo {title} {{Applications of Finite Element
  Methods in reactor mathematics. Numerical solution of the neutron transport
  equation}},\ }in\ \href {https://www.osti.gov/biblio/4237082} {\emph
  {\bibinfo {booktitle} {Argonne National Lab Report}}},\ \bibinfo {series and
  number} {\bibinfo {number} {ANL-8126}}\ (\bibinfo {year} {1974})\BibitemShut
  {NoStop}%
\bibitem [{\citenamefont {Cockburn}(2003)}]{Cockburn2003Nov}%
  \BibitemOpen
  \bibfield  {author} {\bibinfo {author} {\bibfnamefont {B.}~\bibnamefont
  {Cockburn}},\ }\bibfield  {title} {\bibinfo {title} {{Discontinuous Galerkin
  methods}},\ }\href {https://doi.org/10.1002/zamm.200310088} {\bibfield
  {journal} {\bibinfo  {journal} {Z. angew. Math. Mech.}\ }\textbf {\bibinfo
  {volume} {83}},\ \bibinfo {pages} {731} (\bibinfo {year} {2003})}\BibitemShut
  {NoStop}%
\bibitem [{\citenamefont {Sauls}\ and\ \citenamefont
  {Eschrig}(2009)}]{Sauls2009}%
  \BibitemOpen
  \bibfield  {author} {\bibinfo {author} {\bibfnamefont {J.~A.}\ \bibnamefont
  {Sauls}}\ and\ \bibinfo {author} {\bibfnamefont {M.}~\bibnamefont
  {Eschrig}},\ }\bibfield  {title} {\bibinfo {title} {Vortices in chiral,
  spin-triplet superconductors and superfluids},\ }\href
  {https://doi.org/10.1088/1367-2630/11/7/075008} {\bibfield  {journal}
  {\bibinfo  {journal} {New Journal of Physics}\ }\textbf {\bibinfo {volume}
  {11}},\ \bibinfo {pages} {075008} (\bibinfo {year} {2009})}\BibitemShut
  {NoStop}%
\bibitem [{\citenamefont {H{\aa}kansson}\ \emph {et~al.}(2015)\citenamefont
  {H{\aa}kansson}, \citenamefont {L{\ifmmode\ddot{o}\else\"{o}\fi}fwander},\
  and\ \citenamefont
  {Fogelstr{\ifmmode\ddot{o}\else\"{o}\fi}m}}]{Hakansson2015Sep}%
  \BibitemOpen
  \bibfield  {author} {\bibinfo {author} {\bibfnamefont {M.}~\bibnamefont
  {H{\aa}kansson}}, \bibinfo {author} {\bibfnamefont {T.}~\bibnamefont
  {L{\ifmmode\ddot{o}\else\"{o}\fi}fwander}},\ and\ \bibinfo {author}
  {\bibfnamefont {M.}~\bibnamefont
  {Fogelstr{\ifmmode\ddot{o}\else\"{o}\fi}m}},\ }\bibfield  {title} {\bibinfo
  {title} {{Spontaneously broken time-reversal symmetry in high-temperature
  superconductors - Nature Physics}},\ }\href
  {https://doi.org/10.1038/nphys3383} {\bibfield  {journal} {\bibinfo
  {journal} {Nat. Phys.}\ }\textbf {\bibinfo {volume} {11}},\ \bibinfo {pages}
  {755} (\bibinfo {year} {2015})}\BibitemShut {NoStop}%
\bibitem [{\citenamefont {Cockburn}\ \emph {et~al.}(2000)\citenamefont
  {Cockburn}, \citenamefont {Karniadakis},\ and\ \citenamefont
  {Shu}}]{Cockburn2000}%
  \BibitemOpen
  \bibfield  {author} {\bibinfo {author} {\bibfnamefont {B.}~\bibnamefont
  {Cockburn}}, \bibinfo {author} {\bibfnamefont {G.~E.}\ \bibnamefont
  {Karniadakis}},\ and\ \bibinfo {author} {\bibfnamefont {C.-W.}\ \bibnamefont
  {Shu}},\ }\bibfield  {title} {\bibinfo {title} {{The Development of
  Discontinuous Galerkin Methods}},\ }in\ \href
  {https://doi.org/10.1007/978-3-642-59721-3_1} {\emph {\bibinfo {booktitle}
  {{Discontinuous Galerkin Methods}}}}\ (\bibinfo  {publisher} {Springer},\
  \bibinfo {address} {Berlin, Germany},\ \bibinfo {year} {2000})\ pp.\ \bibinfo
  {pages} {3--50}\BibitemShut {NoStop}%
\bibitem [{\citenamefont {Amundsen}\ and\ \citenamefont
  {Linder}(2016)}]{Amundsen2016Mar}%
  \BibitemOpen
  \bibfield  {author} {\bibinfo {author} {\bibfnamefont {M.}~\bibnamefont
  {Amundsen}}\ and\ \bibinfo {author} {\bibfnamefont {J.}~\bibnamefont
  {Linder}},\ }\bibfield  {title} {\bibinfo {title} {{General solution of 2D
  and 3D superconducting quasiclassical systems: coalescing vortices and
  nanoisland geometries}},\ }\href {https://doi.org/10.1038/srep22765}
  {\bibfield  {journal} {\bibinfo  {journal} {Sci. Rep.}\ }\textbf {\bibinfo
  {volume} {6}},\ \bibinfo {pages} {1} (\bibinfo {year} {2016})}\BibitemShut
  {NoStop}%
\bibitem [{\citenamefont {Lahabi}\ \emph {et~al.}(2017)\citenamefont {Lahabi},
  \citenamefont {Amundsen}, \citenamefont {Ouassou}, \citenamefont {Beukers},
  \citenamefont {Pleijster}, \citenamefont {Linder}, \citenamefont {Alkemade},\
  and\ \citenamefont {Aarts}}]{Lahabi2017Dec}%
  \BibitemOpen
  \bibfield  {author} {\bibinfo {author} {\bibfnamefont {K.}~\bibnamefont
  {Lahabi}}, \bibinfo {author} {\bibfnamefont {M.}~\bibnamefont {Amundsen}},
  \bibinfo {author} {\bibfnamefont {J.~A.}\ \bibnamefont {Ouassou}}, \bibinfo
  {author} {\bibfnamefont {E.}~\bibnamefont {Beukers}}, \bibinfo {author}
  {\bibfnamefont {M.}~\bibnamefont {Pleijster}}, \bibinfo {author}
  {\bibfnamefont {J.}~\bibnamefont {Linder}}, \bibinfo {author} {\bibfnamefont
  {P.}~\bibnamefont {Alkemade}},\ and\ \bibinfo {author} {\bibfnamefont
  {J.}~\bibnamefont {Aarts}},\ }\bibfield  {title} {\bibinfo {title}
  {{Controlling supercurrents and their spatial distribution in
  ferromagnets}},\ }\href {https://doi.org/10.1038/s41467-017-02236-2}
  {\bibfield  {journal} {\bibinfo  {journal} {Nat. Commun.}\ }\textbf {\bibinfo
  {volume} {8}},\ \bibinfo {pages} {1} (\bibinfo {year} {2017})}\BibitemShut
  {NoStop}%
\bibitem [{\citenamefont {Amundsen}\ and\ \citenamefont
  {Linder}(2017)}]{Amundsen2017Aug}%
  \BibitemOpen
  \bibfield  {author} {\bibinfo {author} {\bibfnamefont {M.}~\bibnamefont
  {Amundsen}}\ and\ \bibinfo {author} {\bibfnamefont {J.}~\bibnamefont
  {Linder}},\ }\bibfield  {title} {\bibinfo {title} {{Supercurrent vortex
  pinball via a triplet Cooper pair inverse Edelstein effect}},\ }\href
  {https://doi.org/10.1103/PhysRevB.96.064508} {\bibfield  {journal} {\bibinfo
  {journal} {Phys. Rev. B}\ }\textbf {\bibinfo {volume} {96}},\ \bibinfo
  {pages} {064508} (\bibinfo {year} {2017})}\BibitemShut {NoStop}%
\bibitem [{\citenamefont {Nagato}\ \emph {et~al.}(1993)\citenamefont {Nagato},
  \citenamefont {Nagai},\ and\ \citenamefont {Hara}}]{Nagato1993}%
  \BibitemOpen
  \bibfield  {author} {\bibinfo {author} {\bibfnamefont {Y.}~\bibnamefont
  {Nagato}}, \bibinfo {author} {\bibfnamefont {K.}~\bibnamefont {Nagai}},\ and\
  \bibinfo {author} {\bibfnamefont {J.}~\bibnamefont {Hara}},\ }\bibfield
  {title} {\bibinfo {title} {{Theory of the Andreev reflection and the density
  of states in proximity contact normal-superconducting infinite
  double-layer}},\ }\href {https://doi.org/10.1007/BF00682280} {\bibfield
  {journal} {\bibinfo  {journal} {J. Low Temp. Phys.}\ }\textbf {\bibinfo
  {volume} {93}},\ \bibinfo {pages} {33} (\bibinfo {year} {1993})}\BibitemShut
  {NoStop}%
\bibitem [{\citenamefont {Schopohl}\ and\ \citenamefont
  {Maki}(1995)}]{Schopohl1995}%
  \BibitemOpen
  \bibfield  {author} {\bibinfo {author} {\bibfnamefont {N.}~\bibnamefont
  {Schopohl}}\ and\ \bibinfo {author} {\bibfnamefont {K.}~\bibnamefont
  {Maki}},\ }\bibfield  {title} {\bibinfo {title} {{Quasiparticle spectrum
  around a vortex line in a d-wave superconductor}},\ }\href
  {https://doi.org/10.1103/PhysRevB.52.490} {\bibfield  {journal} {\bibinfo
  {journal} {Phys. Rev. B}\ }\textbf {\bibinfo {volume} {52}},\ \bibinfo
  {pages} {490} (\bibinfo {year} {1995})}\BibitemShut {NoStop}%
\bibitem [{\citenamefont {Schopohl}(1998)}]{Schopohl1998}%
  \BibitemOpen
  \bibfield  {author} {\bibinfo {author} {\bibfnamefont {N.}~\bibnamefont
  {Schopohl}},\ }\bibfield  {title} {\bibinfo {title} {{Transformation of the
  Eilenberger Equations of Superconductivity to a Scalar Riccati Equation}},\
  }\href {https://arxiv.org/abs/cond-mat/9804064v1} {\bibfield  {journal}
  {\bibinfo  {journal} {arXiv}\ } (\bibinfo {year} {1998})},\ \Eprint
  {https://arxiv.org/abs/cond-mat/9804064} {cond-mat/9804064} \BibitemShut
  {NoStop}%
\bibitem [{\citenamefont {Shelankov}\ and\ \citenamefont
  {Ozana}(2000)}]{Shelankov2000Mar}%
  \BibitemOpen
  \bibfield  {author} {\bibinfo {author} {\bibfnamefont {A.}~\bibnamefont
  {Shelankov}}\ and\ \bibinfo {author} {\bibfnamefont {M.}~\bibnamefont
  {Ozana}},\ }\bibfield  {title} {\bibinfo {title} {{Quasiclassical theory of
  superconductivity: A multiple-interface geometry}},\ }\href
  {https://doi.org/10.1103/PhysRevB.61.7077} {\bibfield  {journal} {\bibinfo
  {journal} {Phys. Rev. B}\ }\textbf {\bibinfo {volume} {61}},\ \bibinfo
  {pages} {7077} (\bibinfo {year} {2000})}\BibitemShut {NoStop}%
\bibitem [{\citenamefont {Eschrig}(2000)}]{eschrig_distribution_2000}%
  \BibitemOpen
  \bibfield  {author} {\bibinfo {author} {\bibfnamefont {M.}~\bibnamefont
  {Eschrig}},\ }\bibfield  {title} {\bibinfo {title} {Distribution functions in
  nonequilibrium theory of superconductivity and {Andreev} spectroscopy in
  unconventional superconductors},\ }\href
  {https://doi.org/10.1103/PhysRevB.61.9061} {\bibfield  {journal} {\bibinfo
  {journal} {Phys. Rev. B}\ }\textbf {\bibinfo {volume} {61}},\ \bibinfo
  {pages} {9061} (\bibinfo {year} {2000})}\BibitemShut {NoStop}%
\bibitem [{\citenamefont {Eschrig}(2009)}]{Eschrig2009Oct}%
  \BibitemOpen
  \bibfield  {author} {\bibinfo {author} {\bibfnamefont {M.}~\bibnamefont
  {Eschrig}},\ }\bibfield  {title} {\bibinfo {title} {{Scattering problem in
  nonequilibrium quasiclassical theory of metals and superconductors: General
  boundary conditions and applications}},\ }\href
  {https://doi.org/10.1103/PhysRevB.80.134511} {\bibfield  {journal} {\bibinfo
  {journal} {Phys. Rev. B}\ }\textbf {\bibinfo {volume} {80}},\ \bibinfo
  {pages} {134511} (\bibinfo {year} {2009})}\BibitemShut {NoStop}%
\bibitem [{\citenamefont {Reid}(1972)}]{Reid1972Jan}%
  \BibitemOpen
  \bibfield  {author} {\bibinfo {author} {\bibfnamefont {W.~T.}\ \bibnamefont
  {Reid}},\ }\href@noop {} {\emph {\bibinfo {title} {{Riccati Differential
  Equations}}}}\ (\bibinfo  {publisher} {Academic Press},\ \bibinfo {address}
  {Cambridge, MA, USA},\ \bibinfo {year} {1972})\BibitemShut {NoStop}%
\bibitem [{\citenamefont {Abrikosov}\ \emph {et~al.}(1975)\citenamefont
  {Abrikosov}, \citenamefont {Gorkov},\ and\ \citenamefont
  {Dzyaloshinski}}]{AGD}%
  \BibitemOpen
  \bibfield  {author} {\bibinfo {author} {\bibfnamefont {A.~A.}\ \bibnamefont
  {Abrikosov}}, \bibinfo {author} {\bibfnamefont {L.~P.}\ \bibnamefont
  {Gorkov}},\ and\ \bibinfo {author} {\bibfnamefont {I.~E.}\ \bibnamefont
  {Dzyaloshinski}},\ }\href@noop {} {\emph {\bibinfo {title} {Methods of
  Quantum Field Theory in Statistical Physics}}}\ (\bibinfo  {publisher} {Dover
  Publications, Inc.},\ \bibinfo {address} {New York},\ \bibinfo {year}
  {1975})\BibitemShut {NoStop}%
\bibitem [{\citenamefont {Ozaki}(2007)}]{Ozaki2007Jan}%
  \BibitemOpen
  \bibfield  {author} {\bibinfo {author} {\bibfnamefont {T.}~\bibnamefont
  {Ozaki}},\ }\bibfield  {title} {\bibinfo {title} {{Continued fraction
  representation of the Fermi-Dirac function for large-scale electronic
  structure calculations}},\ }\href
  {https://doi.org/10.1103/PhysRevB.75.035123} {\bibfield  {journal} {\bibinfo
  {journal} {Phys. Rev. B}\ }\textbf {\bibinfo {volume} {75}},\ \bibinfo
  {pages} {035123} (\bibinfo {year} {2007})}\BibitemShut {NoStop}%
\bibitem [{\citenamefont {Arnold}\ \emph {et~al.}(2002)\citenamefont {Arnold},
  \citenamefont {Brezzi}, \citenamefont {Cockburn},\ and\ \citenamefont
  {Marini}}]{Arnold2002}%
  \BibitemOpen
  \bibfield  {author} {\bibinfo {author} {\bibfnamefont {D.~N.}\ \bibnamefont
  {Arnold}}, \bibinfo {author} {\bibfnamefont {F.}~\bibnamefont {Brezzi}},
  \bibinfo {author} {\bibfnamefont {B.}~\bibnamefont {Cockburn}},\ and\
  \bibinfo {author} {\bibfnamefont {L.~D.}\ \bibnamefont {Marini}},\ }\bibfield
   {title} {\bibinfo {title} {{Unified Analysis of Discontinuous Galerkin
  Methods for Elliptic Problems}},\ }\href
  {https://epubs.siam.org/doi/10.1137/S0036142901384162} {\bibfield  {journal}
  {\bibinfo  {journal} {SIAM J. Numer. Anal.}\ }\textbf {\bibinfo {volume}
  {39}},\ \bibinfo {pages} {1749–1779} (\bibinfo {year} {2002})}\BibitemShut
  {NoStop}%
\bibitem [{\citenamefont {Brezzi}\ \emph {et~al.}(2004)\citenamefont {Brezzi},
  \citenamefont {Marini},\ and\ \citenamefont
  {S{\ifmmode\ddot{u}\else\"{u}\fi}li}}]{Brezzi2004Dec}%
  \BibitemOpen
  \bibfield  {author} {\bibinfo {author} {\bibfnamefont {F.}~\bibnamefont
  {Brezzi}}, \bibinfo {author} {\bibfnamefont {L.~D.}\ \bibnamefont {Marini}},\
  and\ \bibinfo {author} {\bibfnamefont {E.}~\bibnamefont
  {S{\ifmmode\ddot{u}\else\"{u}\fi}li}},\ }\bibfield  {title} {\bibinfo {title}
  {{Discontinuous Galerkin Methods for First-Order Hyperbolic Problems}},\
  }\href {https://doi.org/10.1142/S0218202504003866} {\bibfield  {journal}
  {\bibinfo  {journal} {Math. Models Methods Appl. Sci.}\ }\textbf {\bibinfo
  {volume} {14}},\ \bibinfo {pages} {1893} (\bibinfo {year}
  {2004})}\BibitemShut {NoStop}%
\bibitem [{\citenamefont {Johnson}(2009)}]{Johnson2009Jan}%
  \BibitemOpen
  \bibfield  {author} {\bibinfo {author} {\bibfnamefont {C.}~\bibnamefont
  {Johnson}},\ }\href@noop {} {\emph {\bibinfo {title} {{Numerical Solution of
  Partial Differential Equations by the Finite Element Method (Dover Books on
  Mathematics)}}}}\ (\bibinfo  {publisher} {Dover Publications},\ \bibinfo
  {year} {2009})\BibitemShut {NoStop}%
\bibitem [{\citenamefont {Zaitsev}(1984)}]{Zaitsev:1984}%
  \BibitemOpen
  \bibfield  {author} {\bibinfo {author} {\bibfnamefont {A.~V.}\ \bibnamefont
  {Zaitsev}},\ }\bibfield  {title} {\bibinfo {title} {Quasiclassical equation
  of the theory of superconductivity for contiguous metals and the properties
  of constricted microstructures},\ }\href
  {http://jetp.ras.ru/cgi-bin/e/index/e/59/5/p1015?a=list} {\bibfield
  {journal} {\bibinfo  {journal} {Zh. Eksp. Teor. Fiz.}\ }\textbf {\bibinfo
  {volume} {86}},\ \bibinfo {pages} {1742} (\bibinfo {year} {1984})},\ \bibinfo
  {note} {[Sov. Phys. JETP 59, 1015 (1984)]}\BibitemShut {NoStop}%
\bibitem [{\citenamefont {Millis}\ \emph {et~al.}(1988)\citenamefont {Millis},
  \citenamefont {Rainer},\ and\ \citenamefont {Sauls}}]{Millis:1988}%
  \BibitemOpen
  \bibfield  {author} {\bibinfo {author} {\bibfnamefont {A.}~\bibnamefont
  {Millis}}, \bibinfo {author} {\bibfnamefont {D.}~\bibnamefont {Rainer}},\
  and\ \bibinfo {author} {\bibfnamefont {J.~A.}\ \bibnamefont {Sauls}},\
  }\bibfield  {title} {\bibinfo {title} {Quasiclassical theory of
  superconductivity near magnetically active interfaces},\ }\href
  {https://doi.org/10.1103/PhysRevB.38.4504} {\bibfield  {journal} {\bibinfo
  {journal} {Phys. Rev. B}\ }\textbf {\bibinfo {volume} {38}},\ \bibinfo
  {pages} {4504} (\bibinfo {year} {1988})}\BibitemShut {NoStop}%
\bibitem [{\citenamefont {Fogelstr\"om}(2000)}]{Fogelstrom:2000}%
  \BibitemOpen
  \bibfield  {author} {\bibinfo {author} {\bibfnamefont {M.}~\bibnamefont
  {Fogelstr\"om}},\ }\bibfield  {title} {\bibinfo {title} {Josephson currents
  through spin-active interfaces},\ }\href
  {https://doi.org/10.1103/PhysRevB.62.11812} {\bibfield  {journal} {\bibinfo
  {journal} {Phys. Rev. B}\ }\textbf {\bibinfo {volume} {62}},\ \bibinfo
  {pages} {11812} (\bibinfo {year} {2000})}\BibitemShut {NoStop}%
\bibitem [{\citenamefont {Zhao}\ \emph {et~al.}(2004)\citenamefont {Zhao},
  \citenamefont {L{\ifmmode\ddot{o}\else\"{o}\fi}fwander},\ and\ \citenamefont
  {Sauls}}]{Zhao2004Oct}%
  \BibitemOpen
  \bibfield  {author} {\bibinfo {author} {\bibfnamefont {E.}~\bibnamefont
  {Zhao}}, \bibinfo {author} {\bibfnamefont {T.}~\bibnamefont
  {L{\ifmmode\ddot{o}\else\"{o}\fi}fwander}},\ and\ \bibinfo {author}
  {\bibfnamefont {J.~A.}\ \bibnamefont {Sauls}},\ }\bibfield  {title} {\bibinfo
  {title} {{Nonequilibrium superconductivity near spin-active interfaces}},\
  }\href {https://doi.org/10.1103/PhysRevB.70.134510} {\bibfield  {journal}
  {\bibinfo  {journal} {Phys. Rev. B}\ }\textbf {\bibinfo {volume} {70}},\
  \bibinfo {pages} {134510} (\bibinfo {year} {2004})}\BibitemShut {NoStop}%
\bibitem [{\citenamefont {Seja}\ and\ \citenamefont
  {L{\ifmmode\ddot{o}\else\"{o}\fi}fwander}(2021)}]{Seja2021Sep}%
  \BibitemOpen
  \bibfield  {author} {\bibinfo {author} {\bibfnamefont {K.~M.}\ \bibnamefont
  {Seja}}\ and\ \bibinfo {author} {\bibfnamefont {T.}~\bibnamefont
  {L{\ifmmode\ddot{o}\else\"{o}\fi}fwander}},\ }\bibfield  {title} {\bibinfo
  {title} {{Quasiclassical theory of charge transport across mesoscopic
  normal-metal--superconducting heterostructures with current conservation}},\
  }\href {https://doi.org/10.1103/PhysRevB.104.104502} {\bibfield  {journal}
  {\bibinfo  {journal} {Phys. Rev. B}\ }\textbf {\bibinfo {volume} {104}},\
  \bibinfo {pages} {104502} (\bibinfo {year} {2021})}\BibitemShut {NoStop}%
\bibitem [{\citenamefont {Seja}\ \emph {et~al.}(2022)\citenamefont {Seja},
  \citenamefont {Jacob},\ and\ \citenamefont
  {L{\ifmmode\ddot{o}\else\"{o}\fi}fwander}}]{Seja2022Mar}%
  \BibitemOpen
  \bibfield  {author} {\bibinfo {author} {\bibfnamefont {K.~M.}\ \bibnamefont
  {Seja}}, \bibinfo {author} {\bibfnamefont {L.}~\bibnamefont {Jacob}},\ and\
  \bibinfo {author} {\bibfnamefont {T.}~\bibnamefont
  {L{\ifmmode\ddot{o}\else\"{o}\fi}fwander}},\ }\bibfield  {title} {\bibinfo
  {title} {{Thermopower and thermophase in a $d$-wave superconductor}},\ }\href
  {https://doi.org/10.1103/PhysRevB.105.104506} {\bibfield  {journal} {\bibinfo
   {journal} {Phys. Rev. B}\ }\textbf {\bibinfo {volume} {105}},\ \bibinfo
  {pages} {104506} (\bibinfo {year} {2022})}\BibitemShut {NoStop}%
\bibitem [{\citenamefont {Badia}\ and\ \citenamefont
  {Verdugo}(2020)}]{Badia2020}%
  \BibitemOpen
  \bibfield  {author} {\bibinfo {author} {\bibfnamefont {S.}~\bibnamefont
  {Badia}}\ and\ \bibinfo {author} {\bibfnamefont {F.}~\bibnamefont
  {Verdugo}},\ }\bibfield  {title} {\bibinfo {title} {{Gridap: An extensible
  Finite Element toolbox in Julia}},\ }\href
  {https://doi.org/10.21105/joss.02520} {\bibfield  {journal} {\bibinfo
  {journal} {Journal of Open Source Software}\ }\textbf {\bibinfo {volume}
  {5}},\ \bibinfo {pages} {2520} (\bibinfo {year} {2020})}\BibitemShut
  {NoStop}%
\bibitem [{\citenamefont {Geuzaine}\ and\ \citenamefont
  {Remacle}(2009)}]{Geuzaine2009Sep}%
  \BibitemOpen
  \bibfield  {author} {\bibinfo {author} {\bibfnamefont {C.}~\bibnamefont
  {Geuzaine}}\ and\ \bibinfo {author} {\bibfnamefont {J.-F.}\ \bibnamefont
  {Remacle}},\ }\bibfield  {title} {\bibinfo {title} {{Gmsh: A 3-D finite
  element mesh generator with built-in pre- and post-processing facilities}},\
  }\href {https://doi.org/10.1002/nme.2579} {\bibfield  {journal} {\bibinfo
  {journal} {Int. J. Numer. Methods Eng.}\ }\textbf {\bibinfo {volume} {79}},\
  \bibinfo {pages} {1309} (\bibinfo {year} {2009})}\BibitemShut {NoStop}%
\bibitem [{\citenamefont {Holmvall}(2017)}]{Holmvall2017}%
  \BibitemOpen
  \bibfield  {author} {\bibinfo {author} {\bibfnamefont {P.}~\bibnamefont
  {Holmvall}},\ }\href {https://research.chalmers.se/en/publication/253315}
  {\emph {\bibinfo {title} {Modeling mesoscopic unconventional
  superconductors}}}\ (\bibinfo  {publisher} {Chalmers University of
  Technology},\ \bibinfo {year} {2017})\ \bibinfo {note} {licentiate
  thesis}\BibitemShut {NoStop}%
\bibitem [{\citenamefont {Holmvall}\ \emph {et~al.}(2018)\citenamefont
  {Holmvall}, \citenamefont {Vorontsov}, \citenamefont
  {Fogelstr{\ifmmode\ddot{o}\else\"{o}\fi}m},\ and\ \citenamefont
  {L{\ifmmode\ddot{o}\else\"{o}\fi}fwander}}]{Holmvall2018Jun}%
  \BibitemOpen
  \bibfield  {author} {\bibinfo {author} {\bibfnamefont {P.}~\bibnamefont
  {Holmvall}}, \bibinfo {author} {\bibfnamefont {A.~B.}\ \bibnamefont
  {Vorontsov}}, \bibinfo {author} {\bibfnamefont {M.}~\bibnamefont
  {Fogelstr{\ifmmode\ddot{o}\else\"{o}\fi}m}},\ and\ \bibinfo {author}
  {\bibfnamefont {T.}~\bibnamefont {L{\ifmmode\ddot{o}\else\"{o}\fi}fwander}},\
  }\bibfield  {title} {\bibinfo {title} {{Broken translational symmetry at
  edges of high-temperature superconductors}},\ }\href
  {https://doi.org/10.1038/s41467-018-04531-y} {\bibfield  {journal} {\bibinfo
  {journal} {Nat. Commun.}\ }\textbf {\bibinfo {volume} {9}},\ \bibinfo {pages}
  {1} (\bibinfo {year} {2018})}\BibitemShut {NoStop}%
\bibitem [{\citenamefont {Holmvall}\ \emph {et~al.}(2020)\citenamefont
  {Holmvall}, \citenamefont {Fogelstr{\ifmmode\ddot{o}\else\"{o}\fi}m},
  \citenamefont {L{\ifmmode\ddot{o}\else\"{o}\fi}fwander},\ and\ \citenamefont
  {Vorontsov}}]{Holmvall2020Jan}%
  \BibitemOpen
  \bibfield  {author} {\bibinfo {author} {\bibfnamefont {P.}~\bibnamefont
  {Holmvall}}, \bibinfo {author} {\bibfnamefont {M.}~\bibnamefont
  {Fogelstr{\ifmmode\ddot{o}\else\"{o}\fi}m}}, \bibinfo {author} {\bibfnamefont
  {T.}~\bibnamefont {L{\ifmmode\ddot{o}\else\"{o}\fi}fwander}},\ and\ \bibinfo
  {author} {\bibfnamefont {A.~B.}\ \bibnamefont {Vorontsov}},\ }\bibfield
  {title} {\bibinfo {title} {{Phase crystals}},\ }\href
  {https://doi.org/10.1103/PhysRevResearch.2.013104} {\bibfield  {journal}
  {\bibinfo  {journal} {Phys. Rev. Res.}\ }\textbf {\bibinfo {volume} {2}},\
  \bibinfo {pages} {013104} (\bibinfo {year} {2020})}\BibitemShut {NoStop}%
\bibitem [{\citenamefont {Wennerdal}\ \emph {et~al.}(2020)\citenamefont
  {Wennerdal}, \citenamefont {Ask}, \citenamefont {Holmvall}, \citenamefont
  {L{\ifmmode\ddot{o}\else\"{o}\fi}fwander},\ and\ \citenamefont
  {Fogelstr{\ifmmode\ddot{o}\else\"{o}\fi}m}}]{Wennerdal2020Nov}%
  \BibitemOpen
  \bibfield  {author} {\bibinfo {author} {\bibfnamefont {N.~W.}\ \bibnamefont
  {Wennerdal}}, \bibinfo {author} {\bibfnamefont {A.}~\bibnamefont {Ask}},
  \bibinfo {author} {\bibfnamefont {P.}~\bibnamefont {Holmvall}}, \bibinfo
  {author} {\bibfnamefont {T.}~\bibnamefont
  {L{\ifmmode\ddot{o}\else\"{o}\fi}fwander}},\ and\ \bibinfo {author}
  {\bibfnamefont {M.}~\bibnamefont
  {Fogelstr{\ifmmode\ddot{o}\else\"{o}\fi}m}},\ }\bibfield  {title} {\bibinfo
  {title} {{Breaking time-reversal and translational symmetry at edges of
  $d$-wave superconductors: Microscopic theory and comparison with
  quasiclassical theory}},\ }\href
  {https://doi.org/10.1103/PhysRevResearch.2.043198} {\bibfield  {journal}
  {\bibinfo  {journal} {Phys. Rev. Res.}\ }\textbf {\bibinfo {volume} {2}},\
  \bibinfo {pages} {043198} (\bibinfo {year} {2020})}\BibitemShut {NoStop}%
\bibitem [{\citenamefont {Chakraborty}\ \emph {et~al.}(2022)\citenamefont
  {Chakraborty}, \citenamefont {L{\ifmmode\ddot{o}\else\"{o}\fi}fwander},
  \citenamefont {Fogelstr{\ifmmode\ddot{o}\else\"{o}\fi}m},\ and\ \citenamefont
  {Black-Schaffer}}]{Chakraborty2022Apr}%
  \BibitemOpen
  \bibfield  {author} {\bibinfo {author} {\bibfnamefont {D.}~\bibnamefont
  {Chakraborty}}, \bibinfo {author} {\bibfnamefont {T.}~\bibnamefont
  {L{\ifmmode\ddot{o}\else\"{o}\fi}fwander}}, \bibinfo {author} {\bibfnamefont
  {M.}~\bibnamefont {Fogelstr{\ifmmode\ddot{o}\else\"{o}\fi}m}},\ and\ \bibinfo
  {author} {\bibfnamefont {A.~M.}\ \bibnamefont {Black-Schaffer}},\ }\bibfield
  {title} {\bibinfo {title} {{Disorder-robust phase crystal in high-temperature
  superconductors stabilized by strong correlations}},\ }\href
  {https://doi.org/10.1038/s41535-022-00450-w} {\bibfield  {journal} {\bibinfo
  {journal} {npj Quantum Mater.}\ }\textbf {\bibinfo {volume} {7}},\ \bibinfo
  {pages} {1} (\bibinfo {year} {2022})}\BibitemShut {NoStop}%
\bibitem [{\citenamefont {Trabaldo}\ \emph
  {et~al.}(2020{\natexlab{a}})\citenamefont {Trabaldo}, \citenamefont
  {Ruffieux}, \citenamefont {Andersson}, \citenamefont {Arpaia}, \citenamefont
  {Montemurro}, \citenamefont {Schneiderman}, \citenamefont {Kalaboukhov},
  \citenamefont {Winkler}, \citenamefont {Lombardi},\ and\ \citenamefont
  {Bauch}}]{Trabaldo2020Mar}%
  \BibitemOpen
  \bibfield  {author} {\bibinfo {author} {\bibfnamefont {E.}~\bibnamefont
  {Trabaldo}}, \bibinfo {author} {\bibfnamefont {S.}~\bibnamefont {Ruffieux}},
  \bibinfo {author} {\bibfnamefont {E.}~\bibnamefont {Andersson}}, \bibinfo
  {author} {\bibfnamefont {R.}~\bibnamefont {Arpaia}}, \bibinfo {author}
  {\bibfnamefont {D.}~\bibnamefont {Montemurro}}, \bibinfo {author}
  {\bibfnamefont {J.~F.}\ \bibnamefont {Schneiderman}}, \bibinfo {author}
  {\bibfnamefont {A.}~\bibnamefont {Kalaboukhov}}, \bibinfo {author}
  {\bibfnamefont {D.}~\bibnamefont {Winkler}}, \bibinfo {author} {\bibfnamefont
  {F.}~\bibnamefont {Lombardi}},\ and\ \bibinfo {author} {\bibfnamefont
  {T.}~\bibnamefont {Bauch}},\ }\bibfield  {title} {\bibinfo {title}
  {{Properties of grooved Dayem bridge based YBa2Cu3 O 7 {-} {$\delta$}
  superconducting quantum interference devices and magnetometers}},\ }\href
  {https://doi.org/10.1063/5.0001805} {\bibfield  {journal} {\bibinfo
  {journal} {Appl. Phys. Lett.}\ }\textbf {\bibinfo {volume} {116}},\ \bibinfo
  {pages} {132601} (\bibinfo {year} {2020}{\natexlab{a}})}\BibitemShut
  {NoStop}%
\bibitem [{\citenamefont {Trabaldo}\ \emph
  {et~al.}(2020{\natexlab{b}})\citenamefont {Trabaldo}, \citenamefont
  {Pfeiffer}, \citenamefont {Andersson}, \citenamefont {Chukharkin},
  \citenamefont {Arpaia}, \citenamefont {Montemurro}, \citenamefont
  {Kalaboukhov}, \citenamefont {Winkler}, \citenamefont {Lombardi},\ and\
  \citenamefont {Bauch}}]{Trabaldo2020Jun}%
  \BibitemOpen
  \bibfield  {author} {\bibinfo {author} {\bibfnamefont {E.}~\bibnamefont
  {Trabaldo}}, \bibinfo {author} {\bibfnamefont {C.}~\bibnamefont {Pfeiffer}},
  \bibinfo {author} {\bibfnamefont {E.}~\bibnamefont {Andersson}}, \bibinfo
  {author} {\bibfnamefont {M.}~\bibnamefont {Chukharkin}}, \bibinfo {author}
  {\bibfnamefont {R.}~\bibnamefont {Arpaia}}, \bibinfo {author} {\bibfnamefont
  {D.}~\bibnamefont {Montemurro}}, \bibinfo {author} {\bibfnamefont
  {A.}~\bibnamefont {Kalaboukhov}}, \bibinfo {author} {\bibfnamefont
  {D.}~\bibnamefont {Winkler}}, \bibinfo {author} {\bibfnamefont
  {F.}~\bibnamefont {Lombardi}},\ and\ \bibinfo {author} {\bibfnamefont
  {T.}~\bibnamefont {Bauch}},\ }\bibfield  {title} {\bibinfo {title} {{SQUID
  Magnetometer Based on Grooved Dayem Nanobridges and a Flux Transformer}},\
  }\href {https://doi.org/10.1109/TASC.2020.3004896} {\bibfield  {journal}
  {\bibinfo  {journal} {IEEE Trans. Appl. Supercond.}\ }\textbf {\bibinfo
  {volume} {30}},\ \bibinfo {pages} {1} (\bibinfo {year}
  {2020}{\natexlab{b}})}\BibitemShut {NoStop}%
\bibitem [{\citenamefont {Hu}(1994)}]{Hu1994}%
  \BibitemOpen
  \bibfield  {author} {\bibinfo {author} {\bibfnamefont {C.-R.}\ \bibnamefont
  {Hu}},\ }\bibfield  {title} {\bibinfo {title} {{Midgap surface states as a
  novel signature for
  ${\mathit{d}}_{\mathit{x}\mathit{a}}^{2}$-${\mathit{x}}_{\mathit{b}}^{2}$-wave
  superconductivity}},\ }\href {https://doi.org/10.1103/PhysRevLett.72.1526}
  {\bibfield  {journal} {\bibinfo  {journal} {Phys. Rev. Lett.}\ }\textbf
  {\bibinfo {volume} {72}},\ \bibinfo {pages} {1526} (\bibinfo {year}
  {1994})}\BibitemShut {NoStop}%
\bibitem [{\citenamefont {Kashiwaya}\ and\ \citenamefont
  {Tanaka}(2000)}]{kashiwaya_tunnelling_2000}%
  \BibitemOpen
  \bibfield  {author} {\bibinfo {author} {\bibfnamefont {S.}~\bibnamefont
  {Kashiwaya}}\ and\ \bibinfo {author} {\bibfnamefont {Y.}~\bibnamefont
  {Tanaka}},\ }\bibfield  {title} {\bibinfo {title} {Tunnelling effects on
  surface bound states in unconventional superconductors},\ }\href
  {https://doi.org/10.1088/0034-4885/63/10/202} {\bibfield  {journal} {\bibinfo
   {journal} {Rep. Prog. Phys.}\ }\textbf {\bibinfo {volume} {63}},\ \bibinfo
  {pages} {1641} (\bibinfo {year} {2000})}\BibitemShut {NoStop}%
\bibitem [{\citenamefont {Löfwander}\ \emph {et~al.}(2001)\citenamefont
  {Löfwander}, \citenamefont {Shumeiko},\ and\ \citenamefont
  {Wendin}}]{lofwander_andreev_2001}%
  \BibitemOpen
  \bibfield  {author} {\bibinfo {author} {\bibfnamefont {T.}~\bibnamefont
  {Löfwander}}, \bibinfo {author} {\bibfnamefont {V.~S.}\ \bibnamefont
  {Shumeiko}},\ and\ \bibinfo {author} {\bibfnamefont {G.}~\bibnamefont
  {Wendin}},\ }\bibfield  {title} {\bibinfo {title} {{Andreev} bound states in
  high-{$T_\mathrm{c}$} superconducting junctions},\ }\href
  {https://doi.org/10.1088/0953-2048/14/5/201} {\bibfield  {journal} {\bibinfo
  {journal} {Supercond. Sci. Technol.}\ }\textbf {\bibinfo {volume} {14}},\
  \bibinfo {pages} {R53} (\bibinfo {year} {2001})}\BibitemShut {NoStop}%
\bibitem [{\citenamefont {Poenicke}\ \emph {et~al.}(1999)\citenamefont
  {Poenicke}, \citenamefont {Barash}, \citenamefont {Bruder},\ and\
  \citenamefont {Istyukov}}]{Poenicke1999}%
  \BibitemOpen
  \bibfield  {author} {\bibinfo {author} {\bibfnamefont {A.}~\bibnamefont
  {Poenicke}}, \bibinfo {author} {\bibfnamefont {Y.~S.}\ \bibnamefont
  {Barash}}, \bibinfo {author} {\bibfnamefont {C.}~\bibnamefont {Bruder}},\
  and\ \bibinfo {author} {\bibfnamefont {V.}~\bibnamefont {Istyukov}},\
  }\bibfield  {title} {\bibinfo {title} {{Broadening of Andreev bound states in
  ${d}_{{x}^{2}\ensuremath{-}{y}^{2}}$ superconductors}},\ }\href
  {https://doi.org/10.1103/PhysRevB.59.7102} {\bibfield  {journal} {\bibinfo
  {journal} {Phys. Rev. B}\ }\textbf {\bibinfo {volume} {59}},\ \bibinfo
  {pages} {7102} (\bibinfo {year} {1999})}\BibitemShut {NoStop}%
\bibitem [{\citenamefont {Holmvall}\ \emph {et~al.}(2022)\citenamefont
  {Holmvall}, \citenamefont {Wennerdal}, \citenamefont {H{\aa}kansson},
  \citenamefont {Stadler}, \citenamefont {Shevtsov}, \citenamefont
  {L{\ifmmode\ddot{o}\else\"{o}\fi}fwander},\ and\ \citenamefont
  {Fogelstr{\ifmmode\ddot{o}\else\"{o}\fi}m}}]{Holmvall2022May}%
  \BibitemOpen
  \bibfield  {author} {\bibinfo {author} {\bibfnamefont {P.}~\bibnamefont
  {Holmvall}}, \bibinfo {author} {\bibfnamefont {N.~W.}\ \bibnamefont
  {Wennerdal}}, \bibinfo {author} {\bibfnamefont {M.}~\bibnamefont
  {H{\aa}kansson}}, \bibinfo {author} {\bibfnamefont {P.}~\bibnamefont
  {Stadler}}, \bibinfo {author} {\bibfnamefont {O.}~\bibnamefont {Shevtsov}},
  \bibinfo {author} {\bibfnamefont {T.}~\bibnamefont
  {L{\ifmmode\ddot{o}\else\"{o}\fi}fwander}},\ and\ \bibinfo {author}
  {\bibfnamefont {M.}~\bibnamefont
  {Fogelstr{\ifmmode\ddot{o}\else\"{o}\fi}m}},\ }\bibfield  {title} {\bibinfo
  {title} {{SuperConga: an open-source framework for mesoscopic
  superconductivity}},\ }\bibfield  {journal} {\bibinfo  {journal} {arXiv}\
  }\href {https://doi.org/10.48550/arXiv.2205.15000}
  {10.48550/arXiv.2205.15000} (\bibinfo {year} {2022}),\ \Eprint
  {https://arxiv.org/abs/2205.15000} {2205.15000} \BibitemShut {NoStop}%
\bibitem [{\citenamefont {Hesthaven}\ and\ \citenamefont
  {Warburton}(2008)}]{Hesthaven2008}%
  \BibitemOpen
  \bibfield  {author} {\bibinfo {author} {\bibfnamefont {J.~S.}\ \bibnamefont
  {Hesthaven}}\ and\ \bibinfo {author} {\bibfnamefont {T.}~\bibnamefont
  {Warburton}},\ }\href
  {https://link.springer.com/book/10.1007/978-0-387-72067-8} {\emph {\bibinfo
  {title} {{Nodal Discontinuous Galerkin Methods}}}}\ (\bibinfo  {publisher}
  {Springer},\ \bibinfo {address} {New York, NY, USA},\ \bibinfo {year}
  {2008})\BibitemShut {NoStop}%
\end{thebibliography}%

\end{document}